\documentclass[12pt,a4paper]{article}

\usepackage{epsfig}

\long\def\symbolfootnote[#1]#2{\begingroup%
\def\thefootnote{\fnsymbol{footnote}}\footnote[#1]{#2}\endgroup}

\begin{document}


\noindent
DESY 04-195, CPT-2004/P.107, WUB 04-17\hfill{\tt hep-lat/0411022\_v2}\\
Nov 2004 -- Jan 2005
\vspace{20pt}

\begin{center}
{\LARGE\bf Scaling tests with dynamical overlap\\[2mm]
and rooted staggered fermions}
\end{center}
\vspace{10pt}

\begin{center}
{\large\bf Stephan D\"urr$\,{}^{a,b}$}
\hspace{8pt}{\large and}\hspace{8pt}
{\large\bf Christian Hoelbling$\,{}^{c,d}$}
\\[10pt]
${}^a\,$DESY Zeuthen, Platanenallee 6, D-15738 Zeuthen, Germany\\
${}^b\,$Institut f\"ur theoretische Physik, Universit\"at Bern, CH-3012 Bern,
Switzerland\\
${}^c\,$Centre de Physique Th\'eorique\symbolfootnote[1]%
{Unit{\'e} Mixte de Recherche (UMR 6207) du CNRS et des Universit{\'e}s Aix
Marseille 1, Aix Marseille 2 et sud Toulon-Var, affili{\'e}e {\`a} la FRUMAM.},
Case 907, CNRS Luminy, F-13288 Marseille Cedex 9, France\\
${}^d\,$Bergische Universit\"at Wuppertal, Gaussstr.\ 20, D-42119 Wuppertal,
Germany
\end{center}
\vspace{10pt}

\begin{abstract}
\noindent
We present a scaling analysis in the 1-flavor Schwinger model with
the full overlap and the rooted staggered determinant.
In the latter case the chiral and continuum limit of the scalar
condensate do not commute, while for overlap fermions they do.
For the topological susceptibility a universal continuum limit is suggested,
as is for the partition function and the Leutwyler-Smilga sum rule.
In the heavy-quark force no difference is visible even at finite coupling.
Finally, a direct comparison between the complete overlap and the rooted
staggered determinant yields evidence that their ratio is constant up to
$O(a^2)$ effects.
\end{abstract}

\clearpage


\newcommand{\pad}{\partial}
\newcommand{\pas}{\partial\!\!\!/}
\newcommand{\Dsl}{D\!\!\!\!/\,}
\newcommand{\Psl}{P\!\!\!\!/\;\!}
\newcommand{\psl}{p\hspace{-2mm}/}
\newcommand{\hqu}{\hbar}
\newcommand{\ovr}{\over}
\newcommand{\til}{\tilde}
\newcommand{\pri}{^\prime}
\renewcommand{\dag}{^\dagger}
\newcommand{\<}{\langle}
\renewcommand{\>}{\rangle}
\newcommand{\gaf}{\gamma_5}
\newcommand{\lap}{\triangle}
\newcommand{\trc}{\mathrm{tr}}
\newcommand{\Mpi}{M_\pi}
\newcommand{\Fpi}{F_\pi}

\newcommand{\al}{\alpha}
\newcommand{\be}{\beta}
\newcommand{\ga}{\gamma}
\newcommand{\de}{\delta}
\newcommand{\ep}{\epsilon}
\newcommand{\ve}{\varepsilon}
\newcommand{\ze}{\zeta}
\newcommand{\et}{\eta}
\renewcommand{\th}{\theta}
\newcommand{\vt}{\vartheta}
\newcommand{\io}{\iota}
\newcommand{\ka}{\kappa}
\newcommand{\la}{\lambda}
\newcommand{\rh}{\rho}
\newcommand{\vr}{\varrho}
\newcommand{\si}{\sigma}
\newcommand{\ta}{\tau}
\newcommand{\ph}{\phi}
\newcommand{\vp}{\varphi}
\newcommand{\ch}{\chi}
\newcommand{\ps}{\psi}
\newcommand{\om}{\omega}

\newcommand{\psb}{\overline{\psi}}
\newcommand{\etb}{\overline{\eta}}
\newcommand{\psd}{\psi^{\dagger}}
\newcommand{\etd}{\eta^{\dagger}}
\newcommand{\beq}{\begin{equation}}

\newcommand{\eeq}{\end{equation}}
\newcommand{\bdm}{\begin{displaymath}}
\newcommand{\edm}{\end{displaymath}}
\newcommand{\bea}{\begin{eqnarray}}
\newcommand{\eea}{\end{eqnarray}}

\newcommand{\mr}{\mathrm}
\newcommand{\mb}{\mathbf}
\newcommand{\Nf}{{N_{\!f}}}
\newcommand{\Nc}{{N_{\!c}}}
\newcommand{\ri}{\mr{i}}
\newcommand{\DW}{D_\mr{W}}
\newcommand{\DN}{D_\mr{N}}
\newcommand{\MeV}{\,\mr{MeV}}
\newcommand{\GeV}{\,\mr{GeV}}
\newcommand{\fm}{\,\mr{fm}}


\hyphenation{topo-lo-gi-cal simu-la-tion theo-re-ti-cal mini-mum con-tinu-um}


\section{Introduction}

An issue which is of both phenomenological and conceptual relevance is whether
it is a valid approach to use the staggered action to study QCD with
$\Nf\!=\!2$ or $\Nf\!=\!2+1$ dynamical quarks.
This is because the staggered Dirac operator $D^\mr{st}$ leads to 4 degenerate
flavors in the continuum, and the simulations are thus performed by taking the
square root of the staggered determinant (plus a quartic root for the dynamical
strange quark).
While the results of such studies look very promising from the phenomenological
viewpoint~\cite{Davies:2003ik,Aubin:2004fs}, a field-theoretic justification of
the rooting procedure might be hard to find.
The origin of the problem is that no one has constructed, so far, a local
operator which, when raised to the fourth power, reproduces $D^\mr{st}$.

It has been shown that the most naive choice results in a non-local operator
\cite{Bunk:2004br,Hart:2004sz}, but of course a more elaborate construction
might settle the issue.

The evidence in favor of the rooting procedure that comes from NLO staggered
chiral perturbation theory~\cite{proroot_sxpt} needs to be backed by numerical
checks of the associated predictions (some are established~\cite{Aubin:2004fs}),
and the analytical thoughts in~\cite{adams_onedimension} involve a number of
simplifications.
It has been shown --~both in 2D and in 4D~-- that staggered eigenvalues on
smooth enough backgrounds form near-degenerate pairs/quadruples which (apart
from a rescaling factor) mimic the (non-degenerate) overlap eigenvalues on the
same configuration~\cite{DuHo, DuHoWe}. 
And the consequence that rooted staggered fermions satisfy an approximate index
theorem and are in the right random-matrix universality class has been
explicitly verified~\cite{Follana:2004sz,WongWolo}.
However, all these pieces are inconclusive, as they do not say how the rooting
procedure should be matched in the \emph{valence\/} sector.

In the absence of a strict analytic argument, only a careful scaling study 
has some conceptual power, albeit an asymmetric one.
If a continuum limit is found with rooted staggered fermions which agrees with
another approach which is considered conceptually proof, nothing firm can be
said (albeit a failure of the staggered framework seems less likely then).
On the other hand, if the continuum limits disagree, the staggered answer would
be in conflict with universality.
In this note we attempt such a scaling study.
The safe approach against which we shall compare is the overlap formulation
\cite{overlap}, which, however, is far more demanding in terms of CPU time.
The theory in which we will work is the massive $\Nf\!=\!1$ Schwinger model,
i.e.\ just QED in 2D.

\bigskip

We define the massless overlap operator as~\cite{overlap}
\beq
D^\mr{ov}=\rh\,\Big(1+\gaf\,\mr{sign}(\gaf D^\mr{W}_{-\rh})\Big)=\rh\,
\Big(1+{D^\mr{W}_{-\rh}\ovr\sqrt{{D^\mr{W}_{-\rho}}\dag D^\mr{W}_{-\rh}}}\Big)
\label{over_zero}
\eeq
with $D^\mr{W}_{-\rh}$ the Wilson operator at negative mass $-\rho$, and
construct the massive overlap via
\beq
D^\mr{ov}_m=(1\!-{m\ovr 2\rho})D^\mr{ov}+m
\;.
\label{over_mass}
\eeq
The massless operator (\ref{over_zero}) satisfies the Ginsparg-Wilson
relation~\cite{Ginsparg:1981bj}
\beq
D\hat\gaf+\gaf D=0,\qquad
\hat\gaf=\gaf(1-{1\ovr\rho}D)
\label{ginspargwilson}
\eeq
which substitutes the continuum chiral symmetry by the lattice chiral symmetry
group~\cite{Luscher:1998pq} 
\beq
\de\ps=\hat\gaf\ps,\qquad\de\psb=\psb\gaf
\label{symm_luscher}
\eeq
which, in turn, excludes additive mass renormalization and prevents operators
in different chiral multiplets from mixing.
On the lattices considered below, the full Dirac matrix may be kept in memory,
and one can use standard linear algebra routines to perform a singular value
decomposition of the shifted Wilson Dirac operator
$D^\mr{W}_{-\rh}\!=\!USV\dag$, where $U, V$ are unitary matrices and $S\!>\!0$
is diagonal.
The massless overlap operator is then simply given by
\beq
D^\mr{ov}\!=\!\rh(1+UV\dag)
\eeq
and it is straightforward to plug in a mass, see (\ref{over_mass}), and call
further library routines to determine the complete eigenvalue spectrum.
In practice, one may prefer to determine the eigensystem of
${D^\mr{W}_{-\rho}}\dag D^\mr{W}_{-\rh}$ or of $\gaf D^\mr{W}_{-\rh}$, but this
does not bring any change in principle.
We use $\rho\!=\!1$, which we checked, following Ref.~\cite{Hernandez:1998et},
is an almost optimal choice with respect to locality for $\be\!\geq\!4$.

\bigskip

The massless staggered operator reads
\begin{equation}
D^\mr{st}=
{1\ovr2}\sum_{\mu}
\et_\mu(x)
\Big(
U_\mu(x)\de_{x+\hat\mu,y}-U_\mu\dag(x\!-\!\hat\mu)\de_{x-\hat\mu,y}
\Big)
\label{stag_zero}
\end{equation}
with $\et_\mu(x)\!=\!(-1)^{\sum_{\nu<\mu}x_\nu}$, and the massive operator is
simply $D^\mr{st}_m\!=\!D^\mr{st}\!+\!m$.
What remains of the continuum $SU(4)_A$ chiral symmetry group in the case of
(\ref{stag_zero}) is the abelian
\beq
\ch(x)\to\exp\Big(\ri\,\th_A(-1)^{\sum_\nu x_\nu}\Big)\ch(x)
\quad,\qquad
\bar\ch(x)\to\bar\ch(x)\exp\Big(-\ri\,\th_A(-1)^{\sum_\nu x_\nu}\Big)
\;,
\label{symm_susskind}
\eeq
which, however, still protects the fermions against additive mass
renormalization.
The parallel transporter $U_\mu(x)$ in (\ref{stag_zero}) may be replaced by a
weighted sum of gauge-covariant paths from $x$ to
$x\!+\!\hat\mu$~\cite{uvfilteredstag}.
As a result, the most unphysical effects in the staggered formulation, the
``taste-changing'' interactions due to highly virtual gluon exchanges
\cite{Lepage:1998vj}, can be considerably reduced.
The reason is a separation of the relevant low-energy modes from the
regularization dependent (and wildly fluctuating) high-energy modes and this is
why we will speak of ``UV-improved'' or ``UV-filtered'' staggered quarks.
For comparison the same modification will be considered in the overlap
case~\cite{uvfilteredover,DuHo,DuHoWe}, too, but there the effect will be much
smaller.

\bigskip

We will be interested in the Schwinger model (SM) with one active flavor, since
only there the rooting issue exists (in 2D the staggered formulation generates
2 flavors in the continuum), but for completeness let us mention the
relationship to QCD both for $\Nf\!=\!1$ and $\Nf\!\geq\!2$.

In the zero temperature SM with $\Nf\!=\!1$ the scalar condensate
$\chi_\mr{sca}\!=\!\<\psb\ps\>$ at zero quark mass follows from the global
axial anomaly and is given by~\cite{Schwinger:1962tp}
\beq
{\chi_\mr{sca}(m\!=\!0)\ovr e}={\exp(\ga)\ovr2\pi^{3/2}}=0.1599\ldots
\label{schwinger}
\eeq
where $\ga\!=\!0.5772...$ is the Euler constant.
Finite temperature effects will reduce $\ch_\mr{sca}$~\cite{Sachs:en}, but no
temperature will be large enough to really make it vanish.
In other words, for $\Nf\!=\!1$ there is no chiral phase transition,
and thus the situation in the SM is analogous to QCD with $\Nf\!=\!1$.

With $\Nf\!\geq\!2$ and a small mass term the \emph{zero temperature\/} SM
shows a vague similarity to QCD \emph{slightly above\/} the phase transition:
The Polyakov loop does not vanish, and the chiral condensate is almost zero;
the system ``tries'' to break the axial flavor symmetry spontaneously, the
spectrum shows a gap between the ``Schwinger particle'' with mass
$e\sqrt{\Nf/\pi}\!+\!O(m)$ (the analog of the $\et'$) and the $\Nf^2\!-\!1$
light ``Quasi-Goldstones'' with mass
$\Mpi\!\sim\!m^{\Nf/(\Nf+1)}$~\cite{Smilga:1995qf}.
The latter get sterile in the chiral limit (as required by Coleman's
theorem~\cite{MerminWagnerHohenbergColeman}), but the important point is that,
as long as one stays away from the chiral limit, these ``pions'' dominate the
correlators between external $P$- and $A_\mu$-sources at sufficiently long
distance.

A peculiarity of any 2D gauge theory is that the fundamental coupling has the
dimension of a mass.
Furthermore, the SM is super-renormalizable and the coupling $e$ does
\emph{not run\/}, hence the theory is not asymptotically free.
We use these features in taking the liberty to set the scale through $e$.
In the compact lattice formulation, the fundamental charge is related to $\be$
via
\beq
\be={1\ovr(ae)^2}
\;,
\label{scale}
\eeq
(below we will set $a\!=\!1$) and this means that the physical scale in the full
theory depends \emph{only\/} on $\be$ and not on the fermion mass $m$.
Of course, other choices would have been possible.
One might set the scale via the measured slope in the short-distance potential
(which is finite in 2D, since $\lim_{r\to0}V(r)/r\!=\!\mr{const}$ in the
continuum with a constant independent of $\Nf$).
In fact, our choice is just this short-distance option, up to cut-off effects.


\section{Eigenvalue spectra}


\begin{table}[b]
\begin{center}
\begin{tabular}{|l|cccccc|}
\hline
geometry&$8\!\times\!8$&$12\!\times\!12$&$16\!\times\!16$&$20\!\times\!20$&
         $24\!\times\!24$&$28\!\times\!28$\\
$\be$   & $0.8$& $1.8$& $3.2$& $5.0$& $7.2$& $9.8$\\
statistics&10,000&10,000&10,000&10,000&10,000&10,000\\
\hline
\end{tabular}
\end{center}
\vspace{-6mm}
\caption{Survey of matched zero temperature lattices, with statistics for both
types of fermions.}
\label{tab:table1}
\begin{center}
\begin{tabular}{|l|cccccccccc|}
\hline
geom.&$ 8\!\times\! 2$&$16\!\times\! 4$&$24\!\times\! 6$&$32\!\times\!8 $&
      $40\!\times\!10$&$48\!\times\!12$&$56\!\times\!14$&
      $64\!\times\!16$&$72\!\times\!18$&$80\!\times\!20$\\
$\be$&  $0.8$ &  $3.2$ &  $7.2$ & $12.8$ & $20.0$
     & $28.8$ & $39.2$ & $51.2$ & $64.8$ & $80.0$\\
over.&$\!10,\!000\!$&$\!10,\!000\!$&$\!10,\!000\!$&$\!10,\!000\!$
     &$\!10,\!000\!$&  --  &  --  &  --  &  --  &  --  \\
stag.&$\!10,\!000\!$&$\!10,\!000\!$&$\!10,\!000\!$&$\!10,\!000\!$
     &$\!10,\!000\!$&$\!10,\!000\!$&$\!10,\!000\!$&$\!10,\!000\!$
     &$\!10,\!000\!$&$\!10,\!000$\\
\hline
\end{tabular}
\end{center}
\vspace{-6mm}
\caption{Survey of matched $T\!=\!L/4$ lattices, with statistics for overlap
and staggered fermions.}
\label{tab:table2}
\end{table}

The plan is to produce a quenched ensemble, to compute all eigenvalues of the
massless $D^\mr{ov}, D^\mr{st}$ on each configuration and to introduce the
dynamical fermions through reweighting~\cite{Lang:1997ib}.
We prepared a variety of lattices with fixed physical volume, both at zero
temperature (i.e.\ with a time-extent $T\!=\!L$ large compared to all
correlation lengths) and at a temperature given via $T\!=\!L/4$, see
Tables~\ref{tab:table1},~\ref{tab:table2} for details.
In 2D standard APE-smearing already involves the full hypercube, and we give
the staple and the original link weight 0.5 each.
Technically, we thus apply a smearing step and evaluate the fermion matrix on
the resulting background, but we consider it a modification of the
\emph{fermion action\/}.
What is peculiar to the choice of setting the scale through $\sqrt{\be}$ is
that our lattices are still (exactly) matched \emph{after\/} reweighting to
$\Nf\!=\!1,2$.

\bigskip

\begin{figure}[t]
\begin{center}
\vspace{-1.3cm}
\hspace{-2cm}\epsfig{file=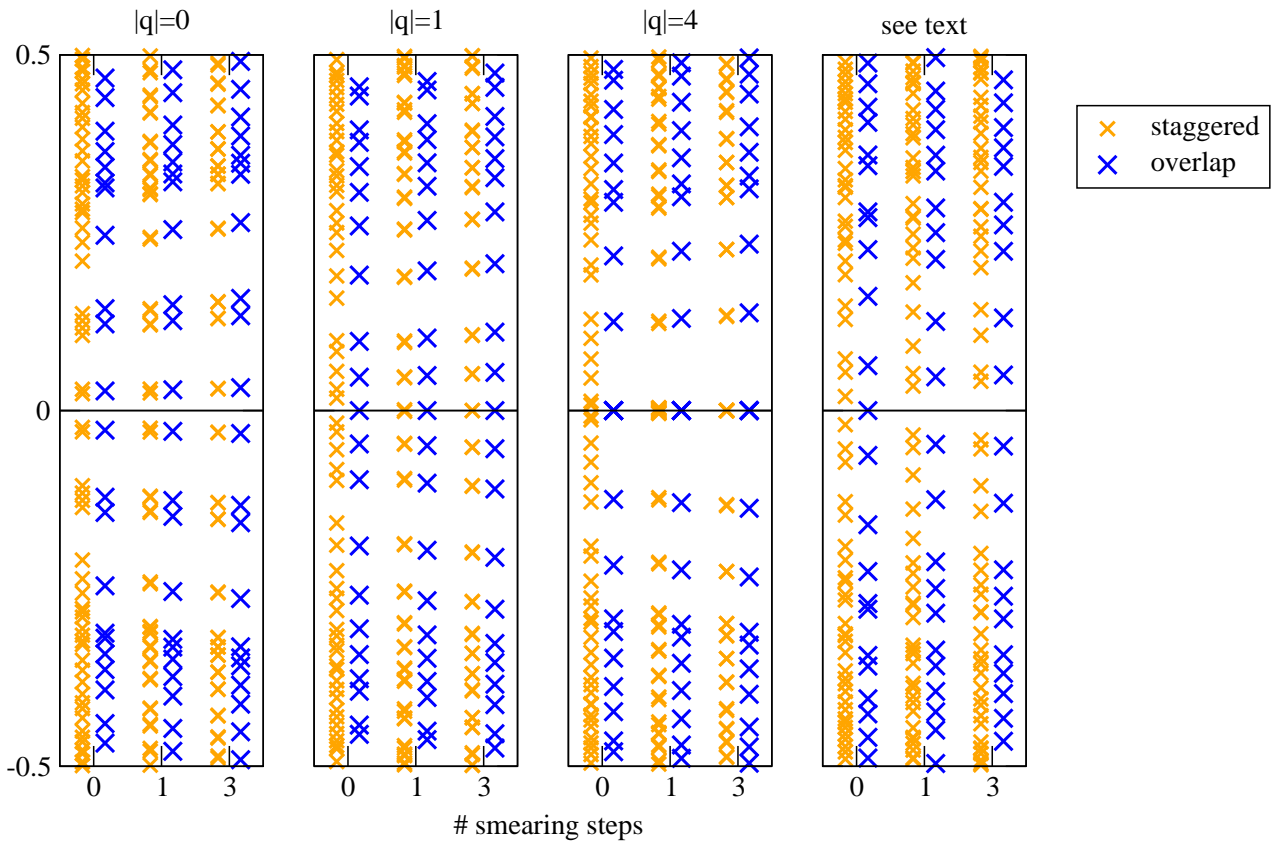,angle=-90,width=17.5cm}
\end{center}
\vspace{-16mm}
\caption{Low-energy spectrum of the unfiltered and two filtered versions of the
Dirac operators $D^\mr{st}$ and $D^\mr{ov}$ [after chiral rotation as in
(\ref{condovereig2})] on three typical configurations at $\be\!=\!7.2$ and on a
selected one where the overlap charge depends on the smearing level (rightmost
panel).}
\label{fig:eigenvalues}
\end{figure}

Fig.~\ref{fig:eigenvalues} shows the low-energy spectrum of $D^\mr{st}$
and the $\hat\la$ of $D^\mr{ov}$ [cf.\ eqn.~(\ref{condovereig2})] on four
selected configurations at $\be\!=\!7.2$.
We define their topological charge as the overlap
index~\cite{Hasenfratz:1998ri,Luscher:1998pq}
\beq
q(U)=\mr{ind}(U)={1\ovr2\rh}\trc(\gaf D^\mr{ov})
\;.
\label{defindex}
\eeq
The first three are typical for topological charge $q=0,1,4$, respectively,
while the last panel shows one of the rare (at $\be\!=\!7.2$) cases where the
overlap charge (\ref{defindex}) depends on the filtering level or, likewise, on
the parameter $\rh$.
Without smearing there is a vague similarity between the staggered and the
overlap spectrum on the $q\!=\!0$ configuration, but not on those with higher
charge.
This holds for fairly smooth gauge fields; the plaquette is $0.927722(54)$ at
$\be\!=\!7.2$.
However, after a single smearing step, the situation improves dramatically.
The staggered eigenvalues form near-degenerate pairs which sit close to an
individual overlap eigenvalue.
In particular the right number of ``would-be'' zero modes separates from the
rest of the spectrum and clusters near the origin, namely $0,2,8$ for
$q\!=\!0,1,4$, respectively.
This is a manifestation of the (approximate) index theorem for staggered
fermions which holds both in 2D~\cite{DuHo} and in
4D~\cite{Follana:2004sz,DuHoWe,WongWolo}.
However, the similarity extends to the higher modes, and one can take the
``fingerprint'' of a configuration likewise with $D^\mr{ov}$ or $D^\mr{st}$,
the two differ just by a trivial rescaling factor and the two-fold degeneracy
in the latter case.
This is the basis of the rooting procedure for staggered fermions, and it is
thus interesting to perform, in a first step, a scaling analysis for some
observables which can be formed from the eigenvalues only.


\section{Scalar condensate}


For overlap fermions, the (bare) scalar condensate is unambiguously defined as
\cite{previousover}
\beq
{\ch_\mr{sca}\ovr e}=
-{\sqrt{\be}\ovr L^2}\<\psb({1\ovr2}+{\gaf\hat\gaf\ovr2})\ps\>=
-{\sqrt{\be}\ovr L^2}\<\psb(1-{1\ovr2\rh}D^\mr{ov})\ps\>
\;.
\label{condover}
\eeq
Denoting the eigenvalues of the massless overlap Dirac operator by $\la$, and
remembering that we work with $\rh\!=\!1$, the reweighted condensate is
\beq
{\ch_\mr{sca}\ovr e}={\sqrt{\be}\ovr L^2}
{\<\det(D^\mr{ov}_m)^\Nf\sum{1-\la/2\ovr(1-m/2)\la+m}\>\ovr
\<\det(D^\mr{ov}_m)^\Nf\>}\;,\qquad
\det(D^\mr{ov}_m)=\prod((1\!-\!{m\ovr 2})\la\!+\!m)
\label{condovereig}
\eeq
where the sum runs over the full spectrum.
These eigenvalues occur either in complex conjugate pairs or as isolated
chiral (doubler) modes at $\lambda\!=\!0\,(2)$.
Finally, one can rewrite (\ref{condovereig}) as
\beq
{\ch_\mr{sca}\ovr e}={\sqrt{\be}\ovr L^2}
{\<\det(D^\mr{ov}_m)^\Nf\sum^{\prime}{1\ovr\hat\la+m}\>\ovr
\<\det(D^\mr{ov}_m)^\Nf\>}\;,\qquad
\hat\la=\Big(\la^{-1}-2^{-1}\Big)^{-1}
\label{condovereig2}
\eeq
where $\hat\la$ is purely imaginary and the primed sum excludes the doubler
modes at $\lambda\!=\!2$.

In the staggered case we follow~\cite{previousstag} and implement the
(bare) 1-flavor condensate through
\beq
{\ch_\mr{sca}\ovr e}=-{\sqrt{\be}\ovr2L^2}\<\bar\ch\ch\>
\;,
\label{condstag}
\eeq
where the purpose of the factor $1/2$ is to compensate the two-fold degeneracy
of the staggered formulation in 2D.
Denoting the eigenvalues of the massless staggered Dirac operator by $\la$
(they show up in complex conjugate pairs with zero real part), the reweighted
condensate is
\beq
{\ch_\mr{sca}\ovr e}={\sqrt{\be}\ovr2L^2}
{\<\det(D^\mr{st}_m)^{\Nf/2}\sum{1\ovr(\la+m)}\>\ovr
\<\det(D^\mr{st}_m)^{\Nf/2}\>}\;,\qquad
\det(D^\mr{st}_m)=\prod(\la\!+\!m)
\;.
\label{condstageig}
\eeq
with the sum and product running over the entire spectrum.

\begin{figure}[t!]
\begin{center}
\epsfig{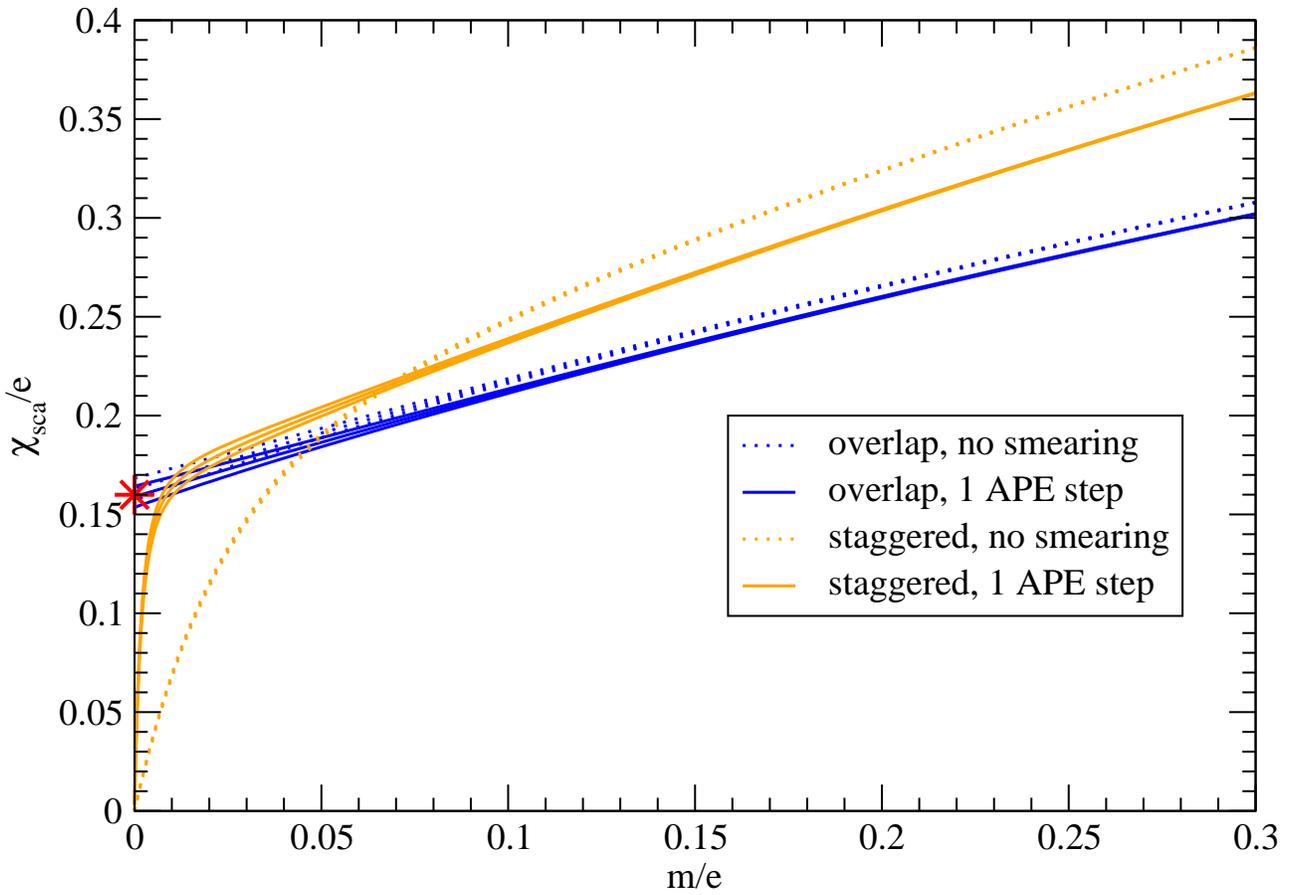}
\end{center}
\vspace{-8mm}
\caption{Bare overlap and rooted staggered $\Nf\!=\!1$ condensate
$\ch_\mr{sca}/e$ at $\be\!=\!7.2$, plotted versus the quark mass. Here and in
subsequent figures, three lines indicate a $1\si$-error band. The overlap
result changes little, if a UV filtered Wilson operator is used instead of an
unfiltered kernel. By contrast, in the staggered case this makes a big
difference -- only the filtered variety shows a clear separation into a regime
(here: $m/e\!>\!0.02$) where the staggered answer is meaningful, and a regime
($m/e\!<\!0.02$) where lattice artefacts overwhelm.}
\label{fig:cond_7.2}
\end{figure}

Finally, there is no renormalization factor to be taken into account in
a 2D theory with cut-off effects of order $O(a^2)$, if one insists on a
massless renormalization scheme.
This follows from the general expansion $Z\!=\!1\!+\!O(e^2)$ which, due to
the dimensionful coupling, reads $Z\!=\!1\!+\!O(a^2e^2)$ and thus tells us
that there is no way to separate a $Z$-factor from intrinsic $O(a^2)$ effects.

\bigskip

An example of the condensates (\ref{condovereig2}, \ref{condstageig}) with
$\Nf\!=\!1$ in one of our zero temperature geometries is shown in
Fig.~\ref{fig:cond_7.2}.
The overlap yields a smooth curve which, for $m\!\to\!0$, is consistent
with the Schwinger value (\ref{schwinger}), marked by an asterisk.
This holds both with a standard Wilson kernel and with the filtered version.
By contrast, the staggered condensate (at any filtering level and any $\be$)
tends to zero, if the chiral limit is performed at fixed lattice spacing, and
taking, in a second step, the continuum limit will not change this, thus
\beq
\lim_{a\to0}\lim_{m\to0}{\ch_\mr{sca}^\mr{st}(m,a)\ovr e}=0
\;.
\eeq
Given the dramatic difference between the two staggered curves in
Fig.~\ref{fig:cond_7.2}, one wonders whether the staggered answer could be
useful, if one uses only the data above some $m_\mr{min}$ to
\emph{extrapolate\/} to the chiral limit.
Obviously, there is no canonical definition of such a $m_\mr{min}$, but the
staggered results in Fig.~\ref{fig:cond_7.2} suggest that it might be less
ambiguous for the filtered operator.
The question can be formalized by asking whether it is possible to reproduce
the Schwinger result (\ref{schwinger}) with rooted staggered fermions, if one
considers the \emph{reverse order of limits\/}, i.e.\ whether
\beq
\lim_{m\to0}\lim_{a\to0}{\ch_\mr{sca}^\mr{st}(m,a)\ovr e}=
{\exp(\ga)\ovr2\pi^{3/2}}
\;.
\label{stag_a_before_m_conjecture}
\eeq
The goal is to show that (only) the second order of limits works with staggered
fermions, while the overlap discretization is correct with either order.

\bigskip

It turns out that eqn.~(\ref{stag_a_before_m_conjecture}) together with our
definitions for $\ch_\mr{sca}^\mr{ov/st}$ does
not make sense, neither for staggered nor for overlap fermions.
This is because the definitions (\ref{condover}) and (\ref{condstag}), when
evaluated with a positive quark-mass, lead to a \emph{logarithmic divergence\/}
in the cut-off.

The logarithmic divergence shows up only in the condensate at finite $m$,
with a coefficient that vanishes as the quark mass tends to zero.
That such a ``soft breaking'' through $m\log(\Lambda^2)$ terms must happen is
obvious from the free case, where elementary manipulations yield
\bea
\<\psb\ps\>\!&\!=\!&\!\int\!d^2p\;{1\ovr \psl-m}\,
=\int\!d^2p\,{m\ovr p^2+m^2}\,
=\int_0^{2\pi}\!d\vp\int_0^\infty\!d\la\;{\la m\ovr\la^2+m^2}
\nonumber
\\
&\!=\!&\!
2\pi\lim_{\Lambda\to\infty}\int_0^{\Lambda}\!d\la\;{\la m\ovr\la^2+m^2}
=\lim_{\Lambda\to\infty} \pi\,m\log\Big({\Lambda^2+m^2\ovr m^2}\Big)
\;.
\label{cond_elementary}
\eea

\begin{figure}[t]
\begin{center}
\epsfig{file=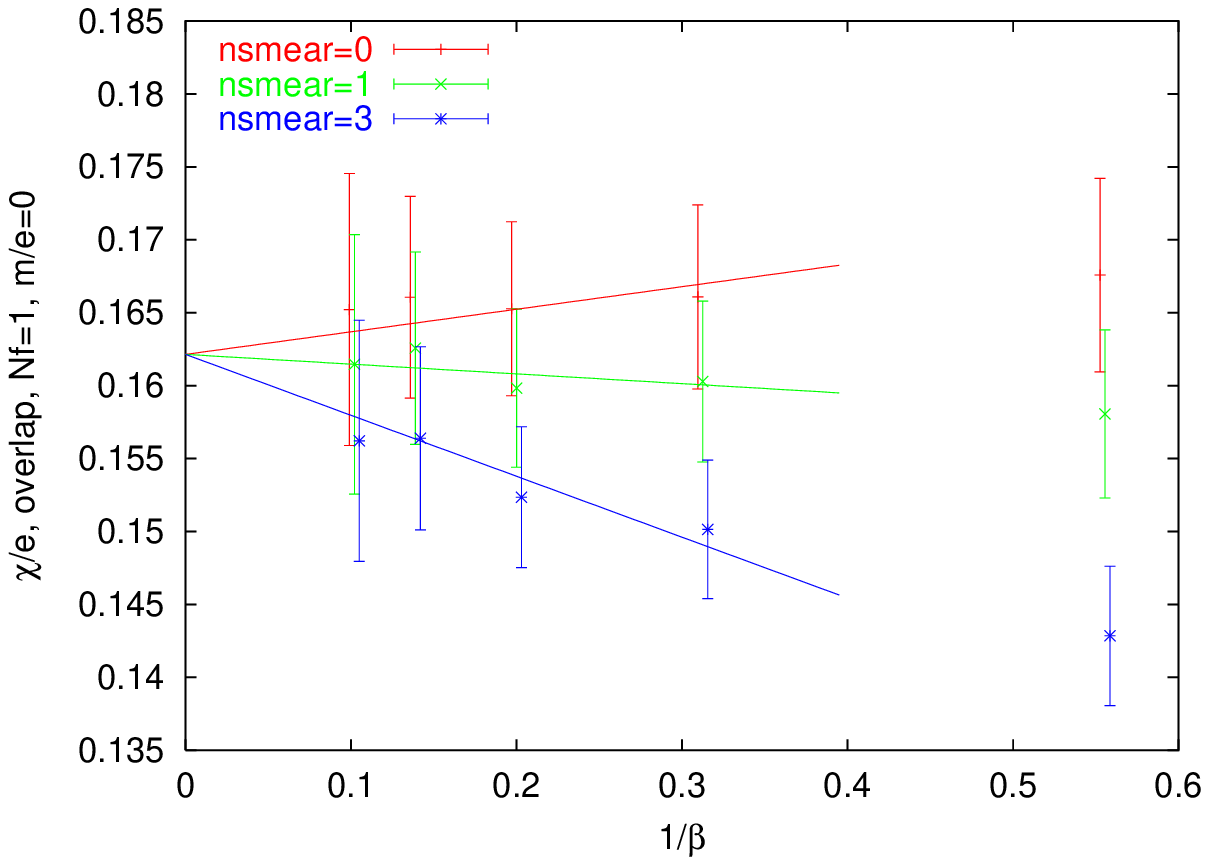,width=8.4cm}
\epsfig{file=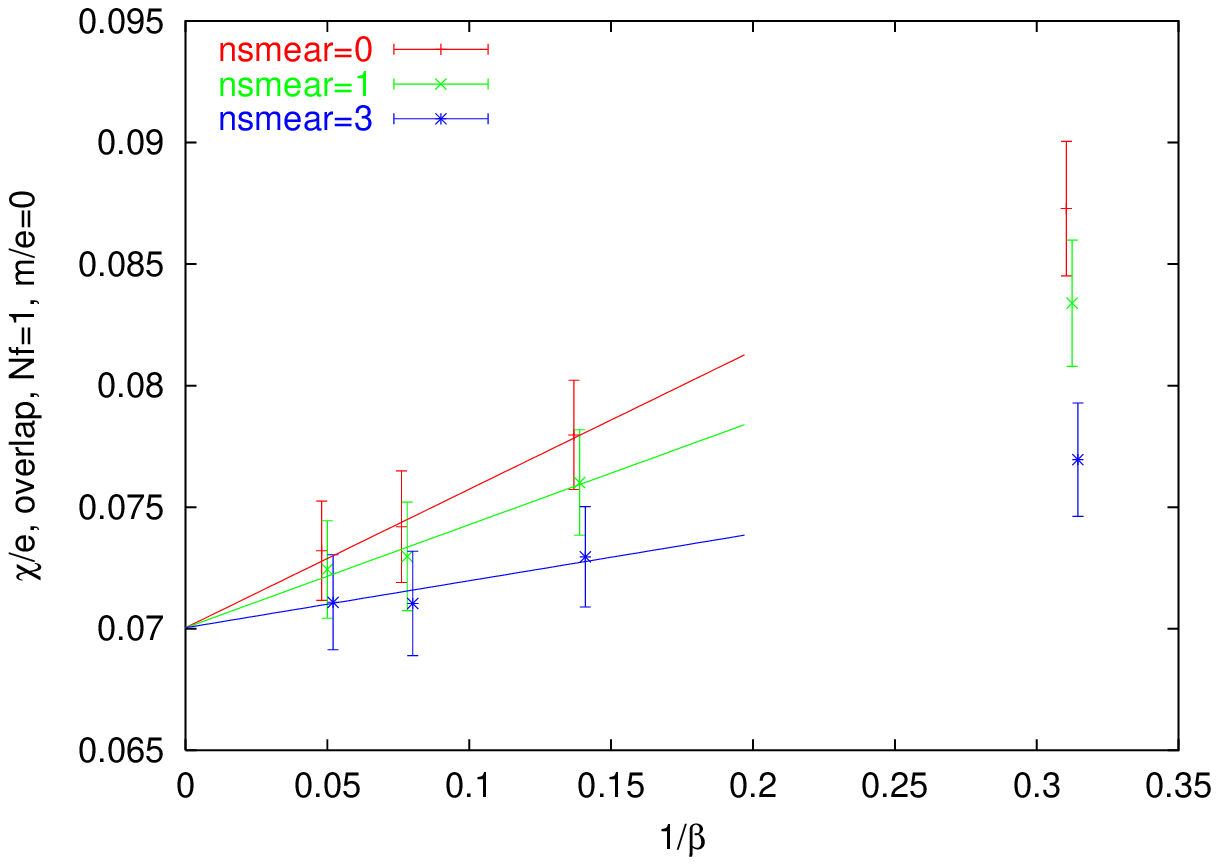,width=8.4cm}
\end{center}
\vspace{-8mm}
\caption{Bare overlap condensate $\ch_\mr{sca}^\mr{ov}/e$ with $m\!=\!0$ at
zero temperature (left) or finite temperature (right) versus $(ae)^2$. The
three filtering levels have a universal continuum limit, and the extrapolation
(\ref{ansatz_zeromass}) works well. The zero temperature result is consistent
with~(\ref{schwinger}).}
\label{fig:cond_zefi_continuum_0.0_over}
\begin{center}
\epsfig{file=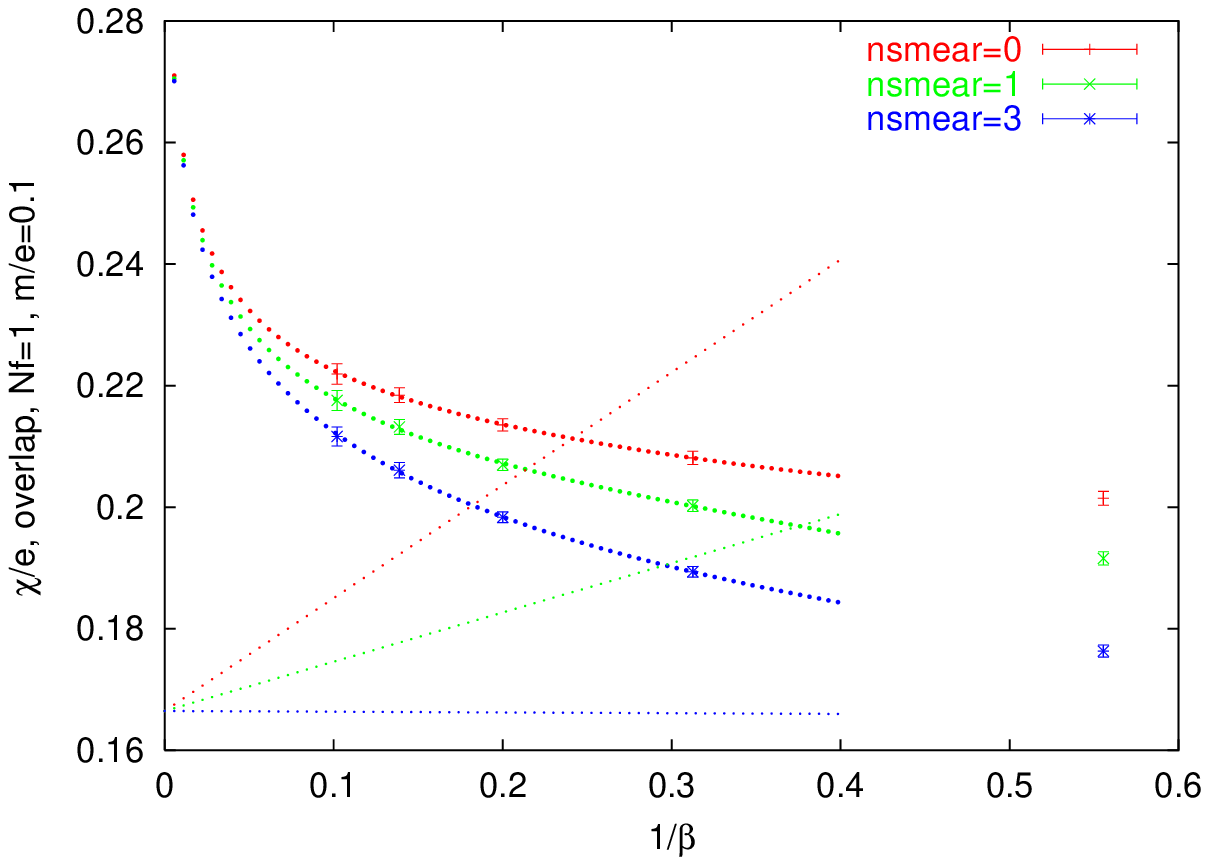,width=8.4cm}
\epsfig{file=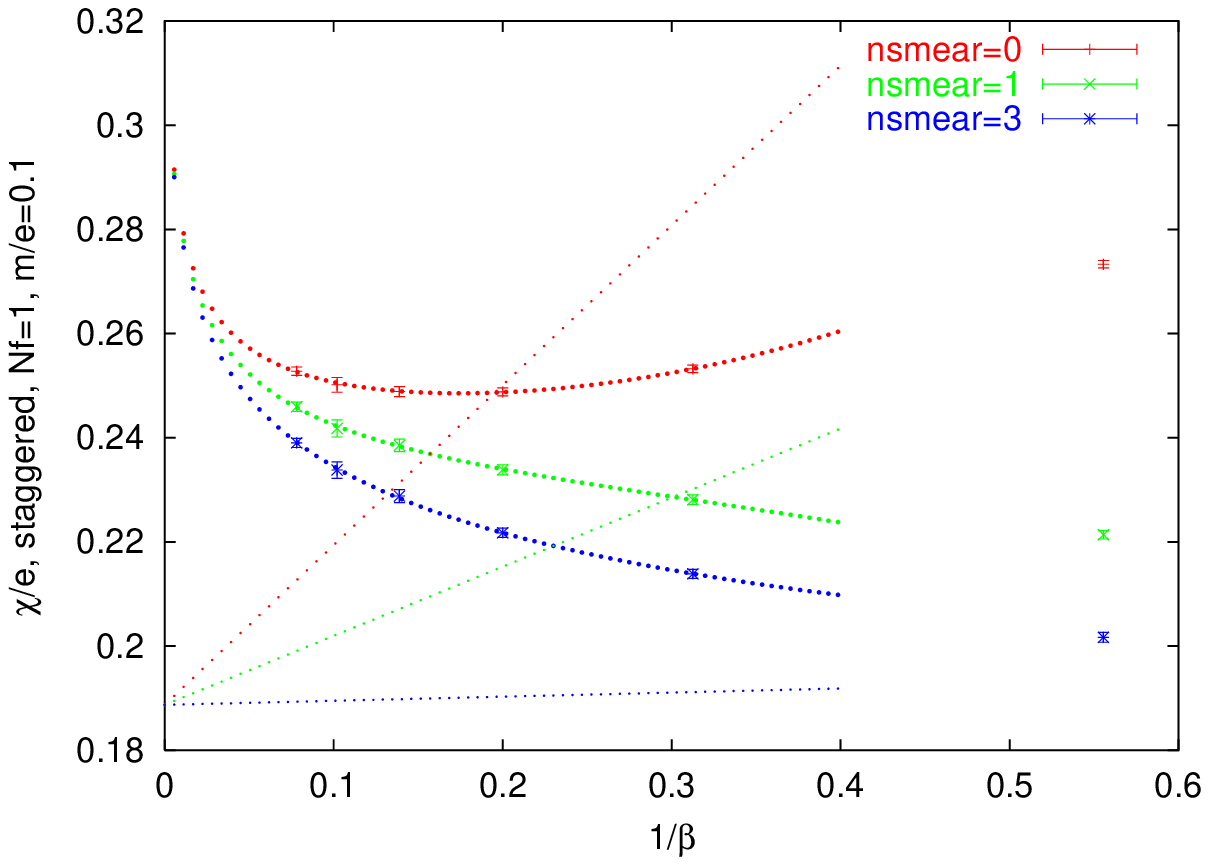,width=8.4cm}
\end{center}
\vspace{-8mm}
\caption{Bare $\Nf\!=\!1$ condensate $\ch_\mr{sca}/e$ with overlap (left) or
rooted staggered (right) fermions at zero temperature and $m/e\!=\!0.1$ versus
$(ae)^2$. The three filtering levels have a common logarithmic
divergence. Adopting the ansatz (\ref{ansatz_finimass}), the divergence may be
subtracted for either discretization, but the so-defined continuum limit is
non-universal.}
\label{fig:cond_squa_continuum_0.1_both}
\end{figure}

\begin{figure}[t]
\begin{center}
\epsfig{file=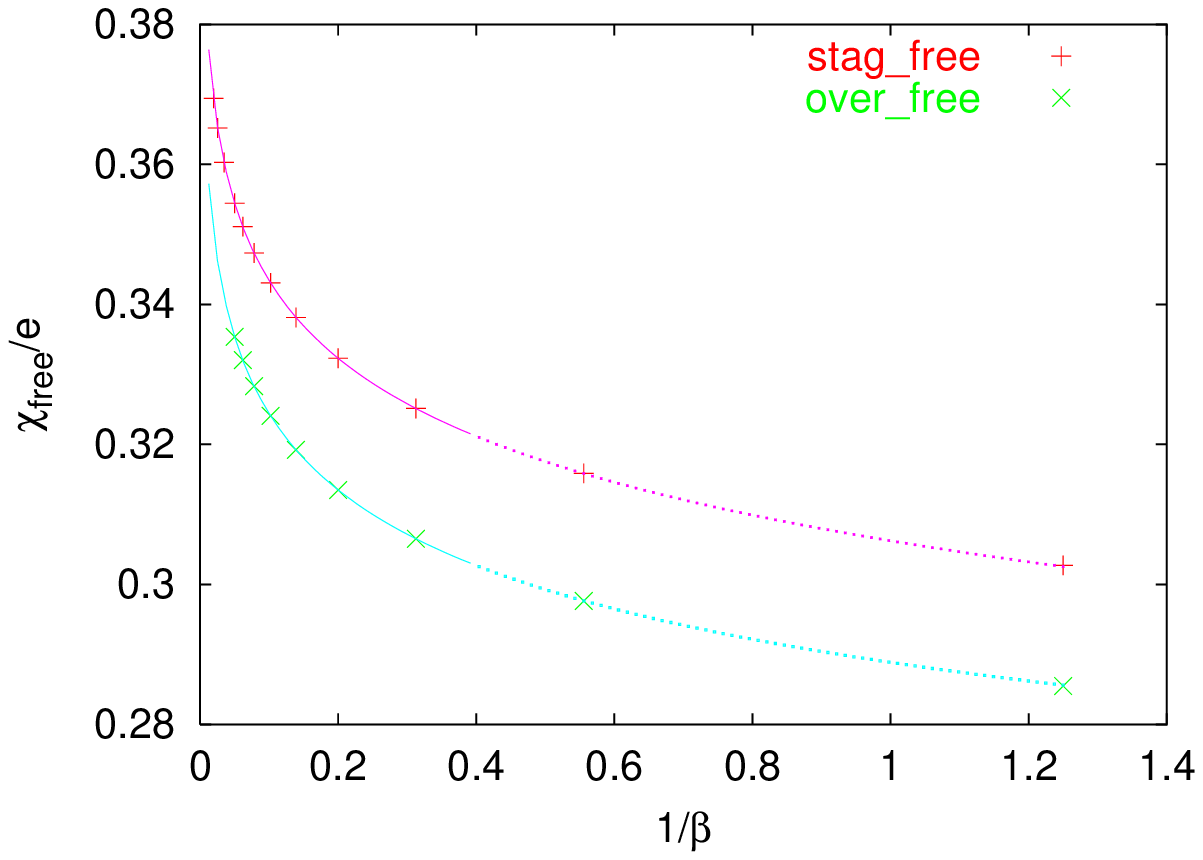,width=8.4cm}
\epsfig{file=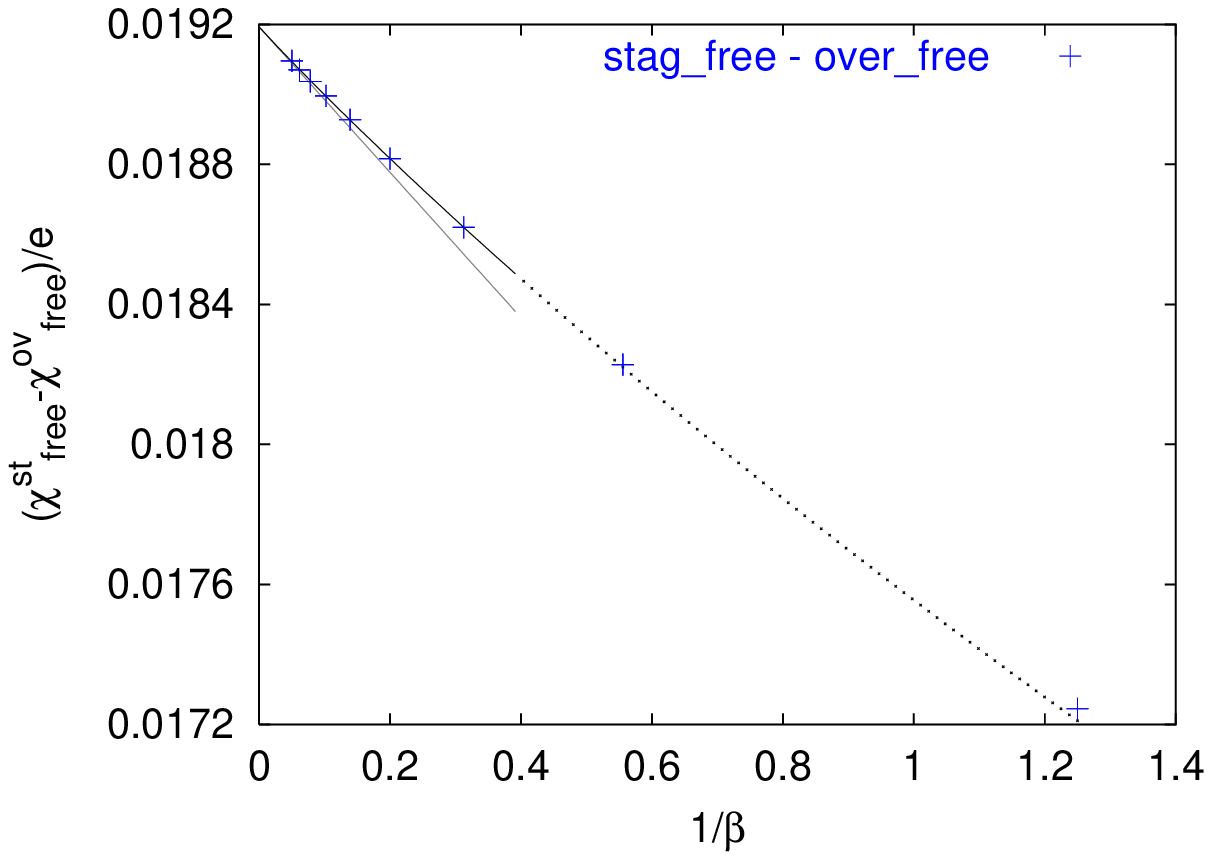,width=8.4cm}
\end{center}
\vspace{-8mm}
\caption{Left: Free condensates versus $1/\be$ (in our zero temperature
geometries, extended towards larger $\be$), at fixed $m/e\!=\!0.1$. The fit to
eqn.\,(\ref{fit_free}) uses the data below 0.4, i.e.\ $L\!\geq\!16$.
Right: Difference of the free staggered and overlap condensate versus $1/\be$.
The data to the left of 0.4 are fitted with a $O(1/\be)\!+\!O(1/\be^{3/2})$
ansatz, shown together with its leading part.}
\label{fig:freecond_0.1}
\end{figure}

Fig.~\ref{fig:cond_zefi_continuum_0.0_over} collects our $\Nf\!=\!1$ overlap
data at zero and finite temperature, evaluated at $m\!=\!0$.
There is no sign of a logarithmic divergence, and we get an acceptable fit if
we use
\beq
{\chi_\mr{sca}(m\!=\!0,1/\be,n)\ovr e}=A+B_n/\be
\label{ansatz_zeromass}
\eeq
with a common $A$ and slope parameters $B_n$ which account for the $O(a^2)$
effects at filtering level $n$.
At zero temperature our continuum result $A\!=\!0.1629(47)$ is compatible with
the Schwinger value~(\ref{schwinger}).
The result $A\!=\!0.0700(16)$ from our thermal geometries indicates that at this
temperature the massless continuum condensate is considerably reduced.
The main lesson is that with overlap fermions one can indeed take the chiral
limit first, letting $a\!\to\!0$ in a second step, and gets the correct answer.

Fig.~\ref{fig:cond_squa_continuum_0.1_both} shows the $\Nf\!=\!1$
zero temperature overlap and staggered condensates, evaluated at $m/e\!=\!0.1$.
Now, the divergence in the cut-off is clearly visible.
Indeed, we get an acceptable result from a correlated fit to all data (in one
formulation) at fixed $m/e$ with the ansatz
\beq
{\chi_\mr{sca}(m/e,1/\be,n)\ovr e}=A+B_n/\be+C\log(1/\be)+
(\de_{n,0}D_0^{(3/2)}\!+\!\de_{n,1}D_1^{(3/2)})/\be^{3/2}+
\de_{n,0}D_0^{(2)}/\be^2
\label{ansatz_finimass}
\eeq
where all coefficients depend on $m/e$, and $A,C$ are common to all filtering
levels, while $B_n$, $D_n^{(3/2,2)}$ depend on the level $n$.
Thus, besides the leading $O(a^2)$ cut-off effects for each filtering level, we
include $O(a^3)$ terms in levels $1,0$, and $O(a^4)$ terms with no smearing.
With rooted staggered quarks everything is analogous, i.e.\ the ansatz
(\ref{ansatz_finimass}) works again.
The fine dotted lines beneath the data indicate what remains if one subtracts
the so-determined logs plus higher order terms and stays with the constant plus
linear part.
For each $m/e\!>\!0$ there is a well-defined continuum limit with such a
procedure, but there is no reason to expect it to be universal.

\bigskip

The difference between the ``minimally subtracted'' overlap and staggered
condensates that survives in the continuum can be traced back to the difference
between the two free condensates.
The latter difference has a finite continuum limit, since each of the free
condensates contains the \emph{same\/} logarithmic divergence.
This point is illustrated in Fig.~\ref{fig:freecond_0.1}.
On the left the free condensates $\ch_\mr{free}^\mr{ov/st}/e$ at fixed physical
quark mass are plotted versus $1/\be$ (here, we refer to
Tab.~\ref{tab:table1}, i.e.\ $\be$ merely encodes for the geometry).
The curves have a common log term; they fit to the form
\beq
{\chi_\mr{free}^\mr{ov/st}(m/e,1/\be)\ovr e}=
A_\mr{free}^\mr{ov/st}(m/e)+
B_\mr{free}^\mr{ov/st}(m/e)/\be+
C_\mr{free}(m/e)\log(1/\be)
\label{fit_free}
\eeq
where $A_\mr{free}, B_\mr{free}$ depend on the discretization, while
$C_\mr{free}$ is universal.
The same conclusion is reached via directly plotting the difference,
as shown on the right; the curvature decreases towards the continuum, and there
is no sign of a remnant log.

The lesson is that one should define the massive continuum condensate in 2D
through
\beq
\ch_\mr{subt}(m/e)=\int_0^\infty\!d\la\;{2m\ovr\la^2+m^2}\;
\Big[\rh(\la)-\rh_\mr{free}(\la)\Big]
\label{cond_continuum}
\eeq
in infinite volume and regard its computation with overlap and staggered
fermions a direct test of universality.
Here, the underlying assumption is that asymptotically the eigenvalue density
$\rh(\la)$ agrees with the free one [from a glimpse at (\ref{cond_elementary}),
one learns that $\rh_\mr{free}(\la)\to\pi\la$].

\begin{figure}[t]
\epsfig{file=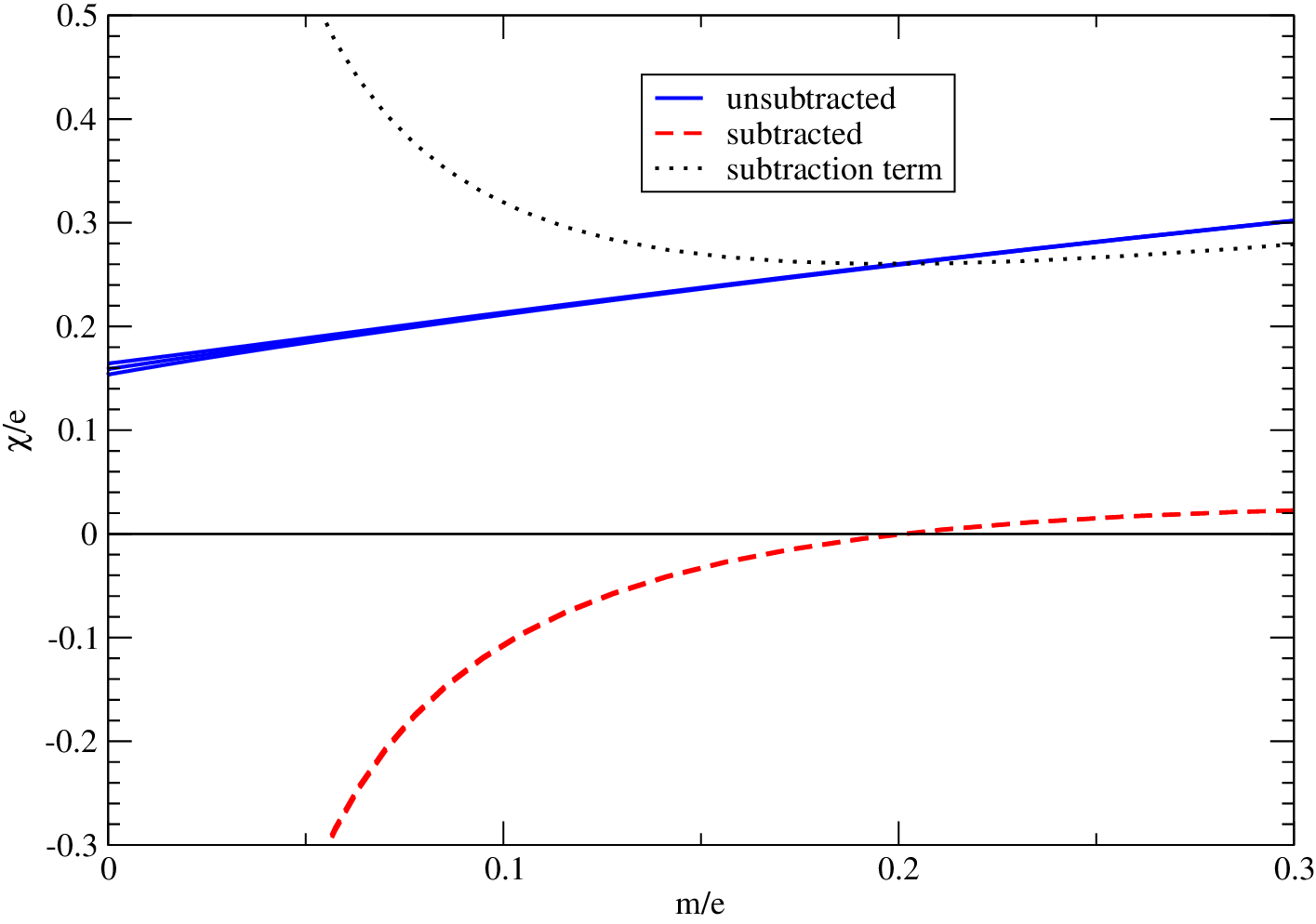,width=8.4cm}\hfill
\epsfig{file=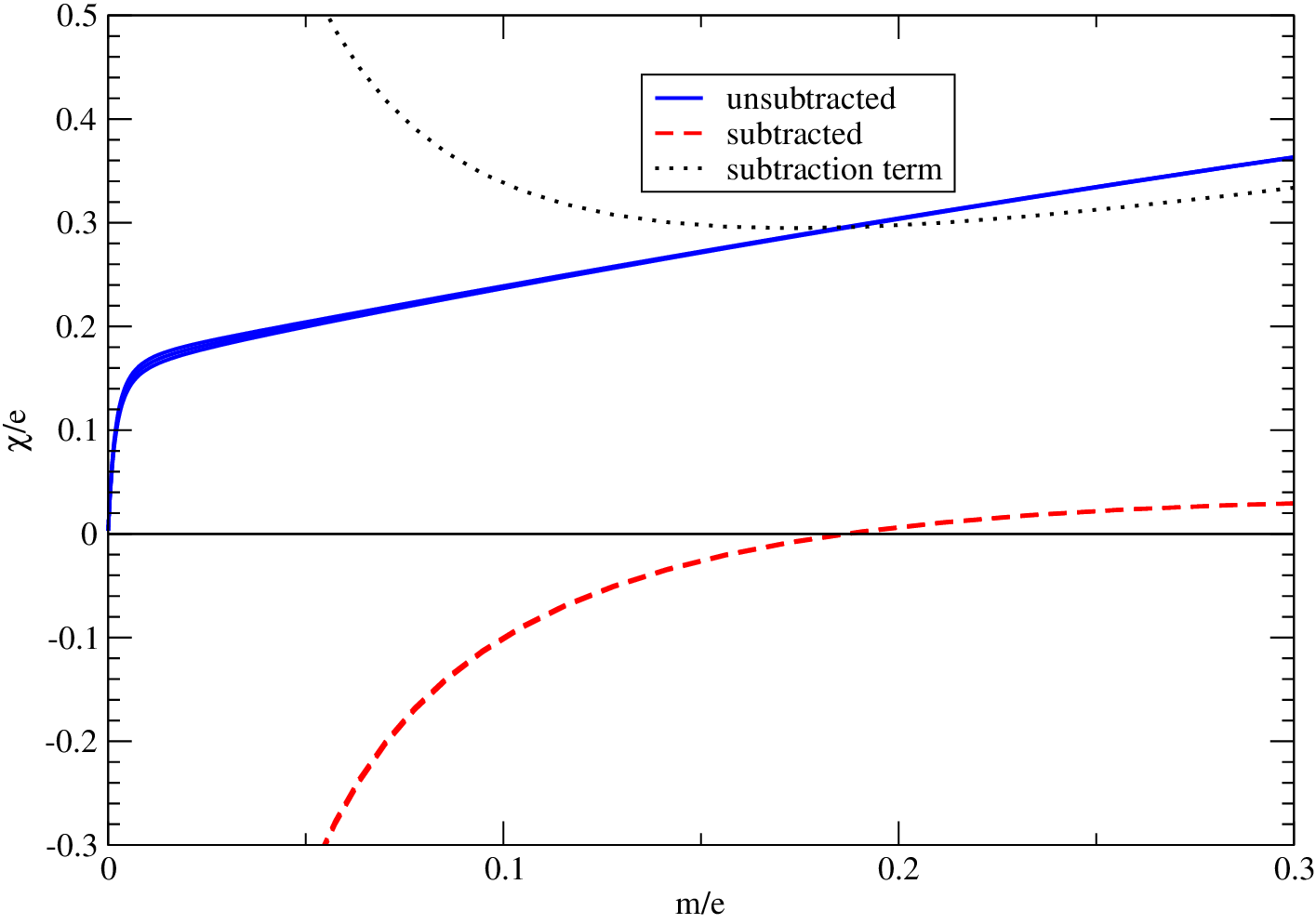,width=8.4cm}
\vspace{-2mm}
\caption{The ``naive'' subtracted condensates (\ref{over_subt_nai},
\ref{stag_subt_nai}) at $\be\!=\!7.2$, together with $\ch_\mr{sca}$ and
$\ch_\mr{free}$ for overlap (left) and staggered (right) fermions. Note the
power-like IR-divergence in $\ch_\mr{subt}^\mr{ov/st}$.}
\label{fig:subt_both_nai}
\vspace{4mm}
\epsfig{file=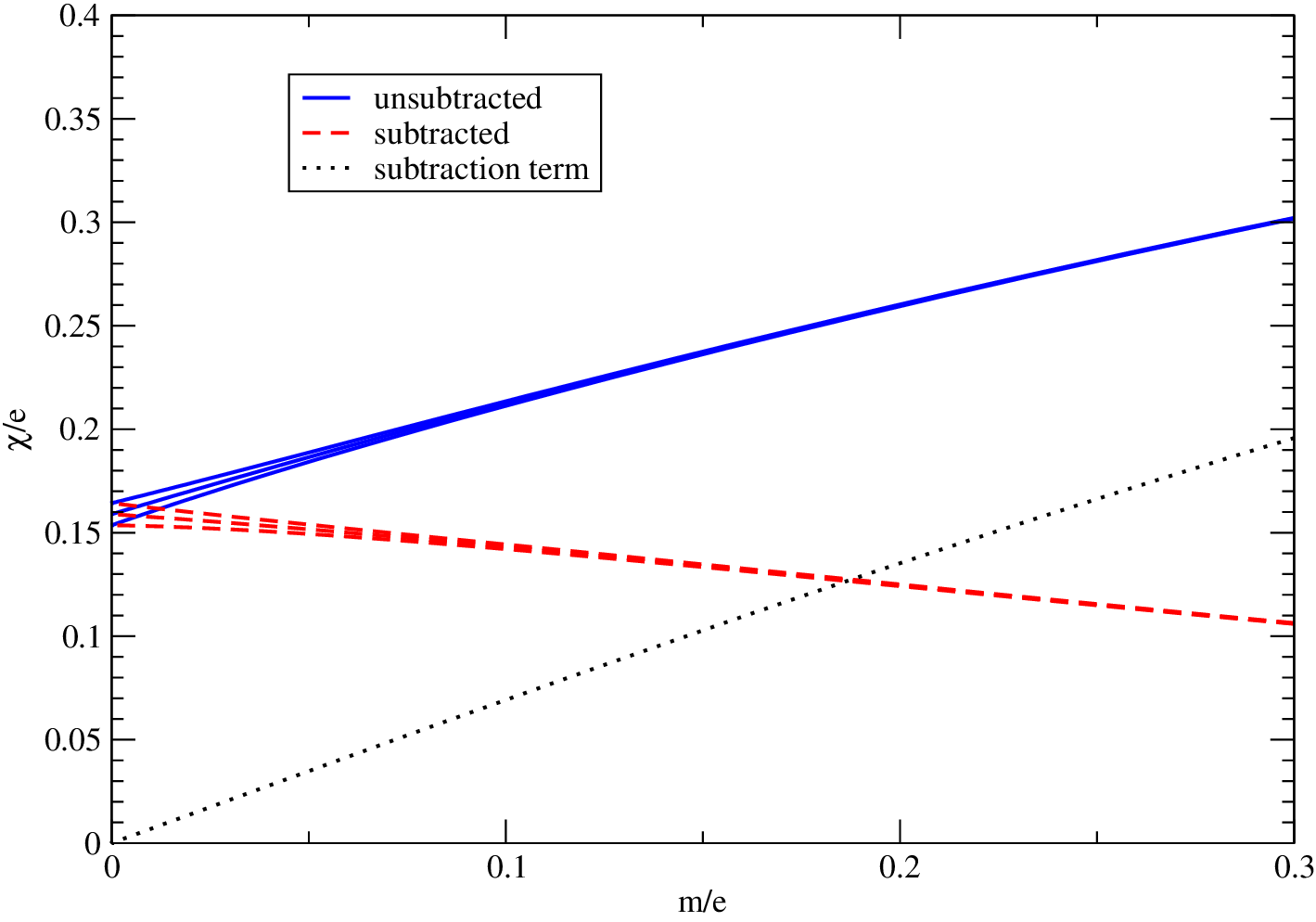,width=8.4cm}\hfill
\epsfig{file=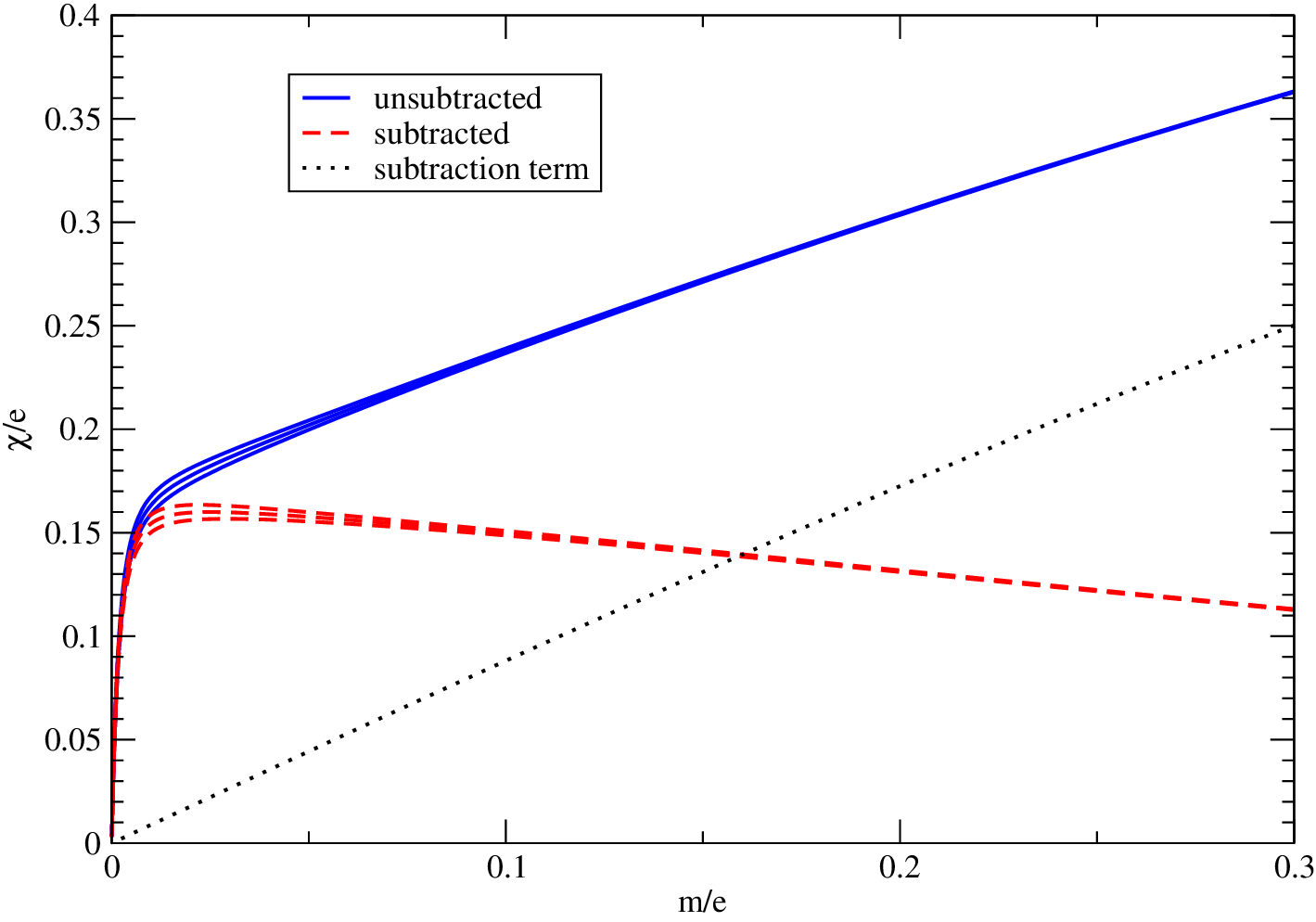,width=8.4cm}
\vspace{-2mm}
\caption{The ``primed'' subtracted condensates (\ref{over_subt_pri},
\ref{stag_subt_pri}) at $\be\!=\!7.2$. No IR-divergence is left.}
\label{fig:subt_both_pri}
\end{figure}

A legitimate strategy to implement (\ref{cond_continuum}) on the lattice is to
first subtract the free case in the same formulation and geometry, followed by
an extrapolation to infinite volume.
The first step may be done analytically, the corresponding expressions read
\bea
\ch_\mr{free}^\mr{ov}(m)&=&{1\ovr L^2}
\sum_{k_1,k_2=0}^{L-1}\mr{Tr}\!\left\{D_\mr{free}^\mr{ov}(m)^{-1}
\Big(1-{1\ovr2\rh}D_\mr{free}^\mr{ov}(0)\Big)\right\}
\label{free_over_nai}
\\
D_\mr{free}^\mr{ov}(m)&=&\Big(\rh-{m\ovr2}\Big)
\left(1+
{D_\mr{free}^\mr{W}(-\rh)\ovr|D_\mr{free}^\mr{W}(-\rho)|}
\right)+m
\nonumber
\\
D_\mr{free}^\mr{W}(M)&=&
2+M-\sum_\mu c_\mu+\ri\sum_\mu s_\mu\ga_\mu
\nonumber
\\
\ch_\mr{free}^\mr{st}(m)&=&{1\ovr L^2}
\sum_{k_1,k_2=0}^{L/2-1}{2m\ovr m^2+s_1^2+s_2^2}
\label{free_stag_nai}
\eea
with $c_\mu=\cos({2\pi\ovr L}k_\mu)$ and $s_\mu=\sin({2\pi\ovr L}k_\mu)$.
In the (``naive'') subtracted condensates
\bea
\ch_\mr{subt}^\mr{ov}(m/e,1/\be,V)&=&
\ch_\mr{sca}^\mr{ov}(m/e,1/\be,V)-\ch_\mr{free}^\mr{ov}(m/e,V)
\label{over_subt_nai}
\\
\ch_\mr{subt}^\mr{st}(m/e,1/\be,V)&=&
\ch_\mr{sca}^\mr{st}(m/e,1/\be,V)-\ch_\mr{free}^\mr{st}(m/e,V)
\label{stag_subt_nai}
\eea
the limit $V\!\to\!\infty$ must be taken before  $m\!\to\!0$, since the free
case contains a massless mode, and this reflects itself in the behavior
$\ch_\mr{subt}^\mr{ov/st}/e\!\to\!-2/(emV)$ for a small mass.

\begin{figure}[t]
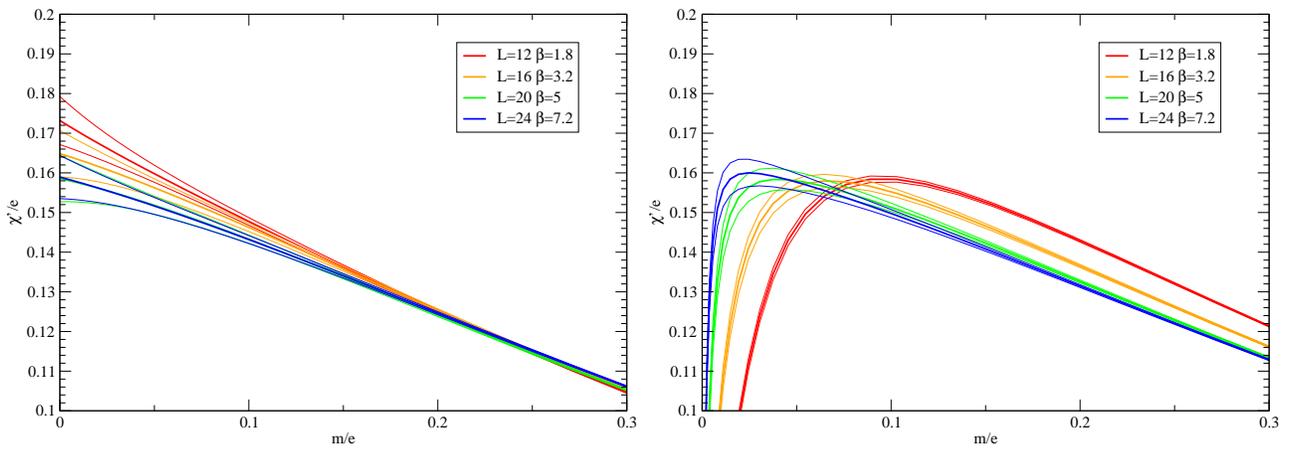

\begin{center}
\epsfig{file=smrw_v2.figs/cond_over_subt1.eps,width=8.4cm}
\epsfig{file=smrw_v2.figs/cond_stag_subt1.eps,width=8.4cm}
\end{center}
\vspace{-8mm}
\caption{The ``primed'' subtracted condensates (\ref{over_subt_pri},
\ref{stag_subt_pri}) for overlap (left) and staggered (right) fermions with 1
smearing step. Note the small scaling violations at any $m/e$ with overlap
fermions, while they get progressively worse for small $m/e$ in the staggered
case.}
\label{fig:subt_pri_allbeta}
\end{figure}

Looking at (\ref{free_over_nai}, \ref{free_stag_nai}) the piece that makes the
chiral limit in a finite volume singular is easily identified, it is the term
with $k_1\!=\!k_2\!=\!0$.
Hence, a better choice is to subtract the free case in the same geometry
\emph{without the non-topological zero-modes\/}.
With
\bea
\ch_\mr{free}^{\prime\;\mr{ov}}(m)&=&{1\ovr L^2}
\sum\nolimits^\prime\mr{Tr}\!\left\{D_\mr{free}^\mr{ov}(m)^{-1}
\Big(1-{1\ovr2\rh}D_\mr{free}^\mr{ov}(0)\Big)\right\}
\label{free_over_pri}
\\
\ch_\mr{free}^{\prime\;\mr{st}}(m)&=&{1\ovr L^2}
\sum\nolimits^\prime{2m\ovr m^2+s_1^2+s_2^2}
\label{free_stag_pri}
\eea
[the primed sum skips the $(0,0)$ contribution, otherwise the summation is as
in (\ref{free_over_nai}, \ref{free_stag_nai})], one may define the ``primed''
subtracted condensates
\bea
\ch_\mr{subt}^{\prime\;\mr{ov}}(m/e,1/\be,V)&=&
\ch_\mr{sca}^\mr{ov}(m/e,1/\be,V)-\ch_\mr{free}^{\prime\;\mr{ov}}(m/e,V)
\label{over_subt_pri}
\\
\ch_\mr{subt}^{\prime\;\mr{st}}(m/e,1/\be,V)&=&
\ch_\mr{sca}^\mr{st}(m/e,1/\be,V)-\ch_\mr{free}^{\prime\;\mr{st}}(m/e,V)
\label{stag_subt_pri}
\eea
that avoid the IR singularity in (\ref{over_subt_nai}, \ref{stag_subt_nai}).
In other words, (\ref{over_subt_pri}, \ref{stag_subt_pri}) is the only lattice
implementation of (\ref{cond_continuum}) which is simultaneously UV and IR
finite and thus permits to skip the limit $V\!\to\!\infty$, because finite
volume corrections are (asymptotically) exponentially small.
Note that this procedure differs from a standard renormalization in QCD.
Since the theory is super-renormalizable, there is no bare parameter to be
adjusted and no counterterm to be added to the Lagrangian, the only option
is a pure vacuum reordering.

The ``naive'' and ``primed'' condensates
(\ref{over_subt_nai}, \ref{stag_subt_nai}) and
(\ref{over_subt_pri}, \ref{stag_subt_pri}) are shown for one coupling in
Fig.~\ref{fig:subt_both_nai} and Fig.~\ref{fig:subt_both_pri}, both for overlap
and staggered quarks.
The former version clearly exhibits the IR divergence
$\ch_\mr{subt}^\mr{ov/st}(m/e\!\to\!0)/e\!\to\!-2/(emV)$, due to the
subtraction term in (\ref{over_subt_nai}, \ref{stag_subt_nai}).
In the latter version, the leading finite-volume effects are tied to the
lightest particle in the \emph{interacting\/} theory, the Schwinger particle
with mass $e/\sqrt{\pi}\!+\!O(m)$, and thus exponentially small.

\begin{figure}[t!]
\epsfig{file=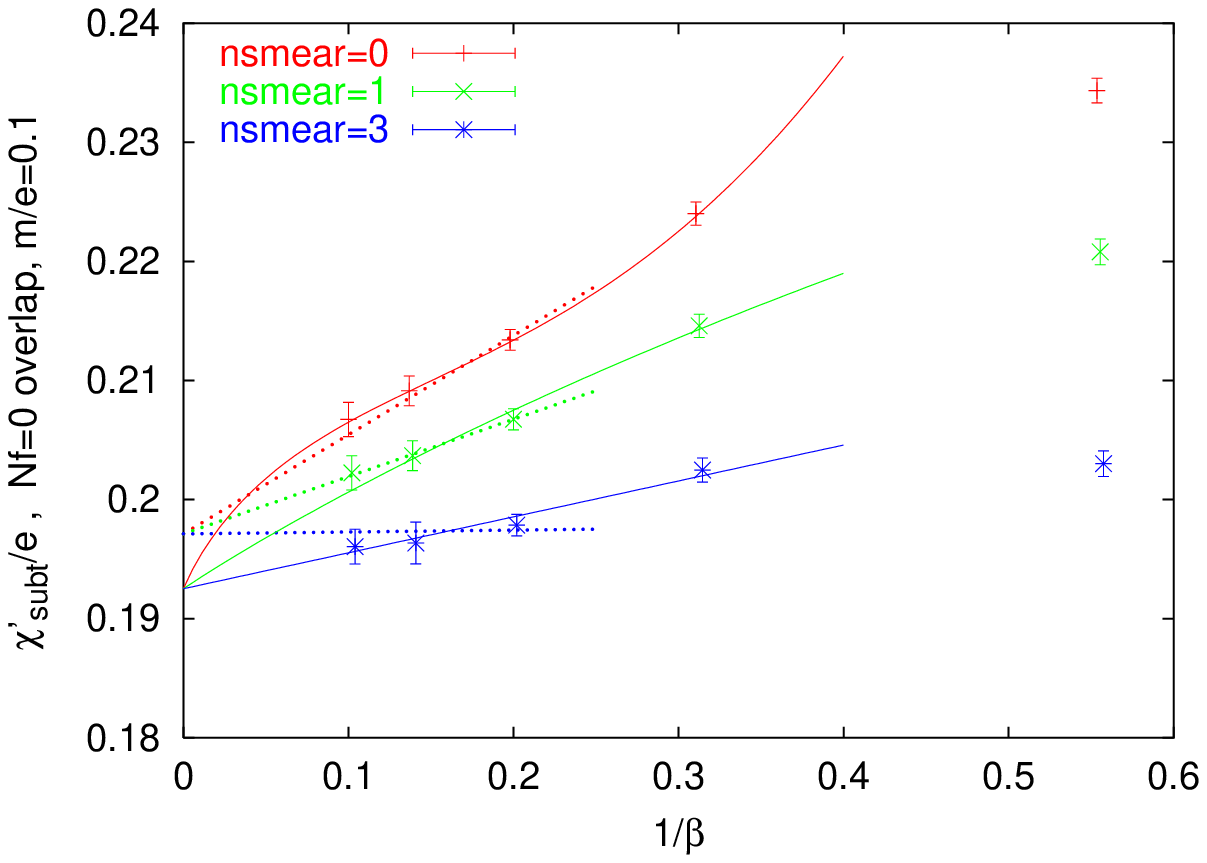,width=8.4cm}
\epsfig{file=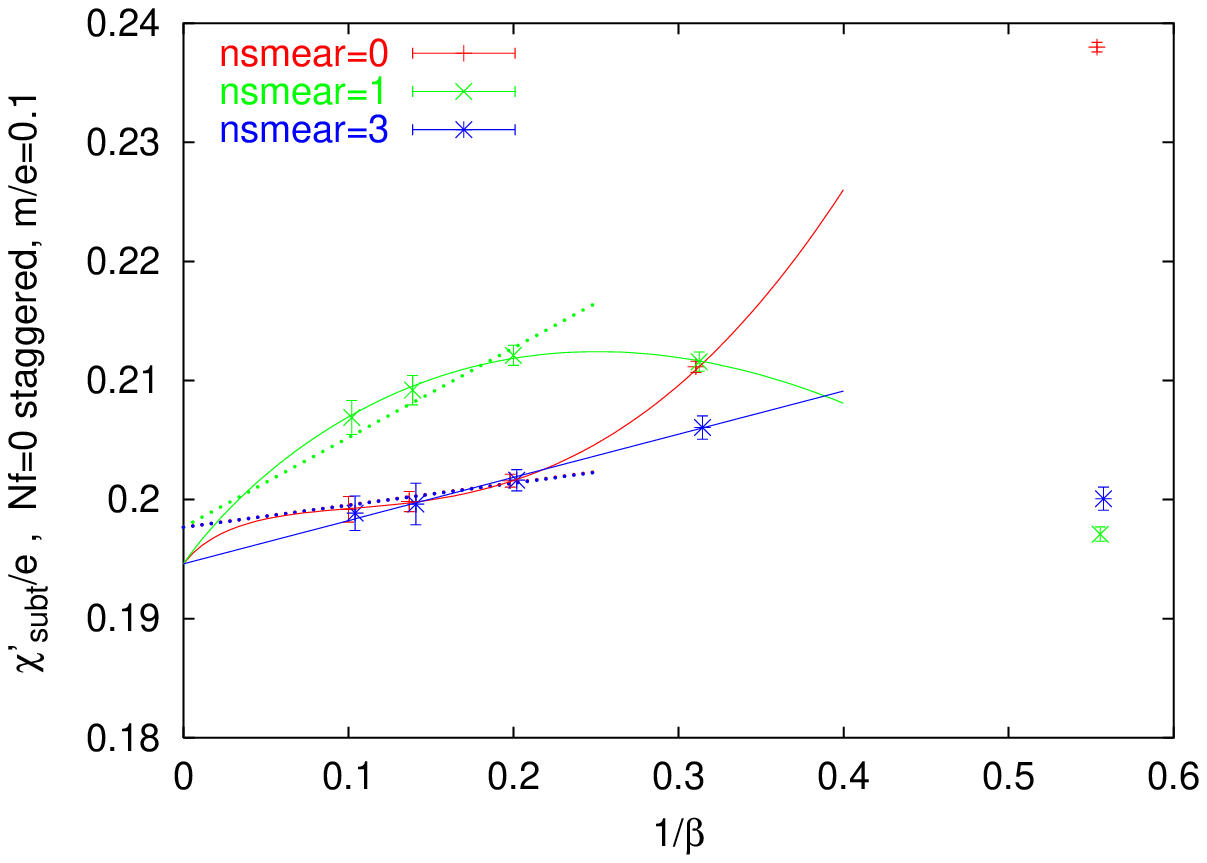,width=8.4cm}
\epsfig{file=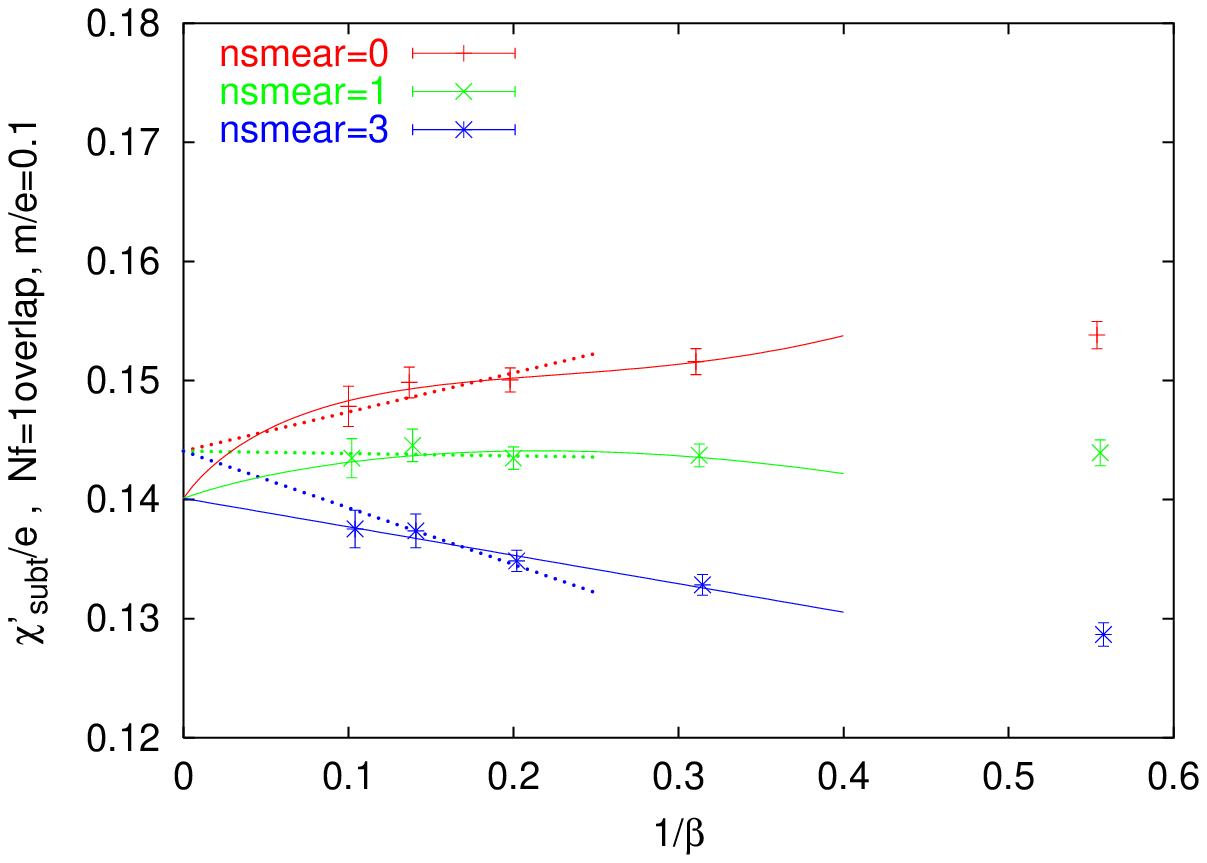,width=8.4cm}
\epsfig{file=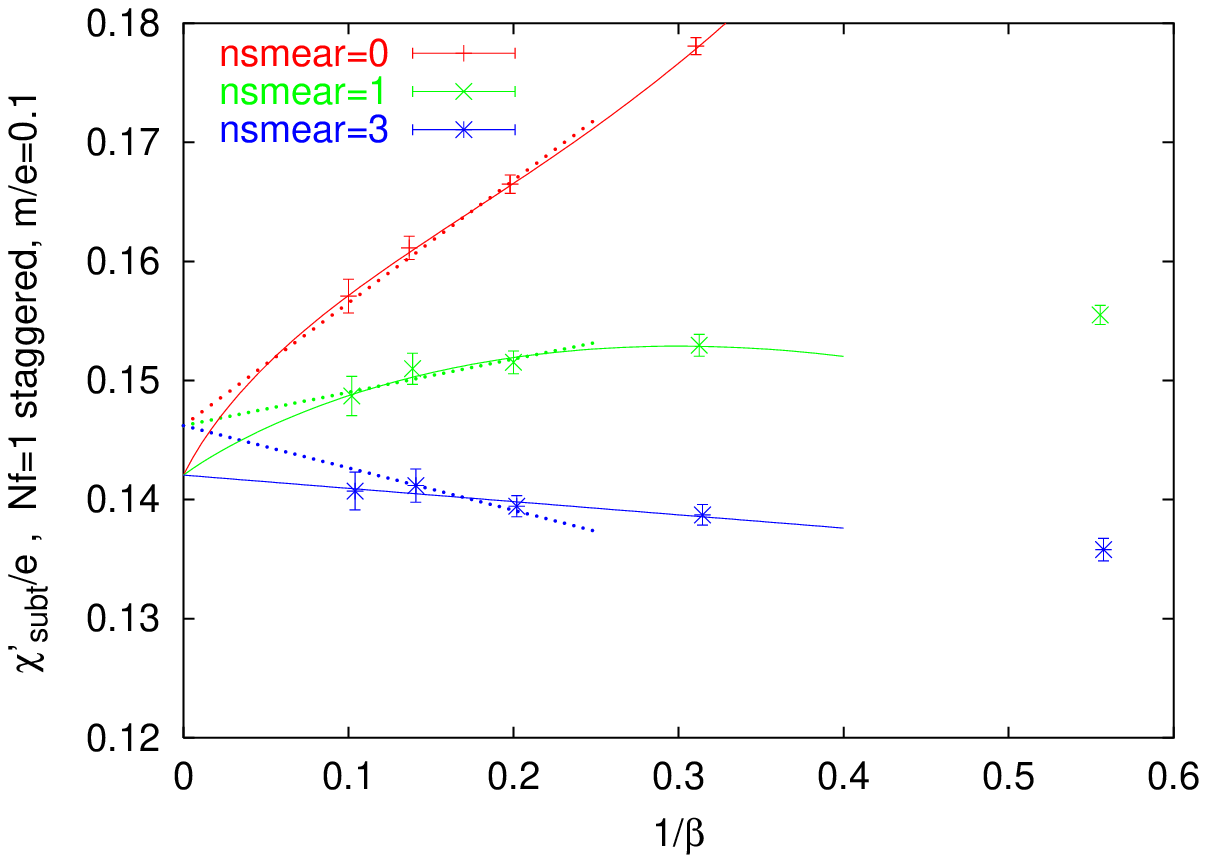,width=8.4cm}
\epsfig{file=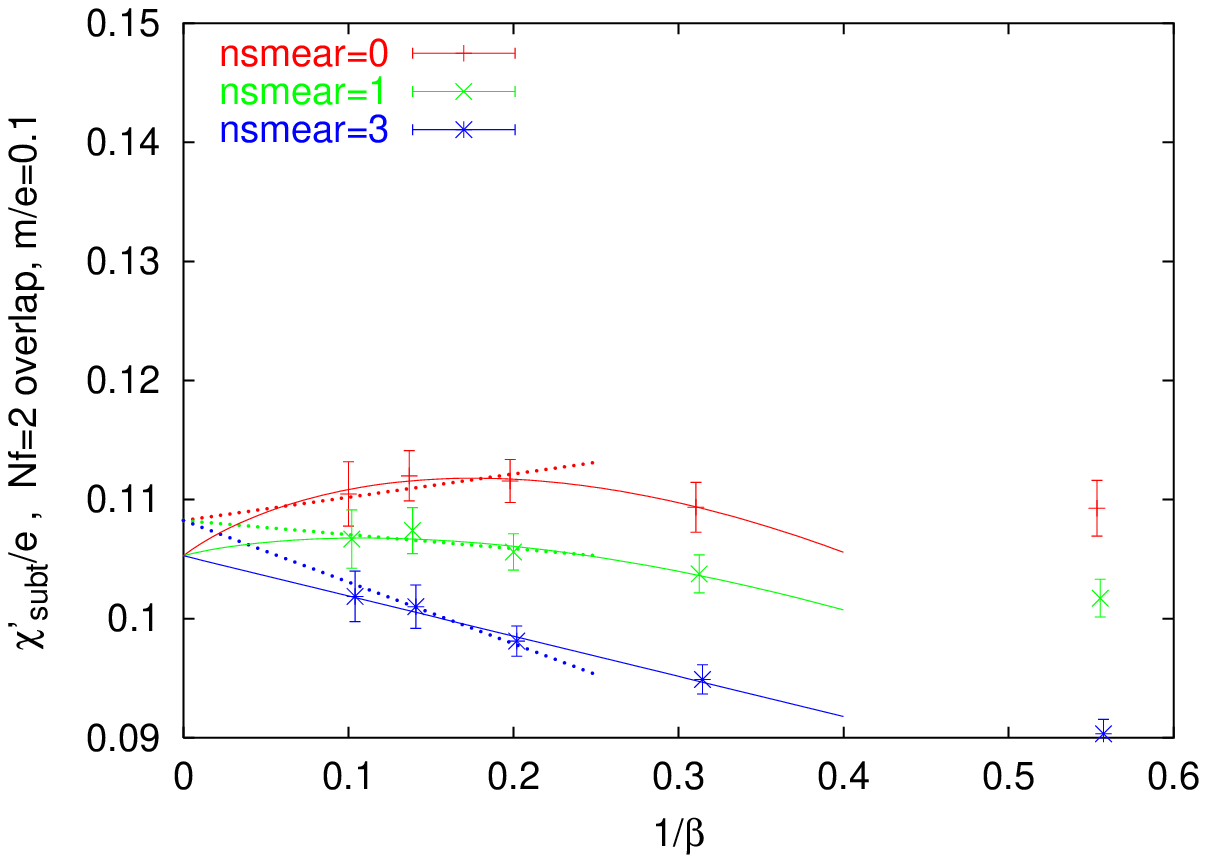,width=8.4cm}
\epsfig{file=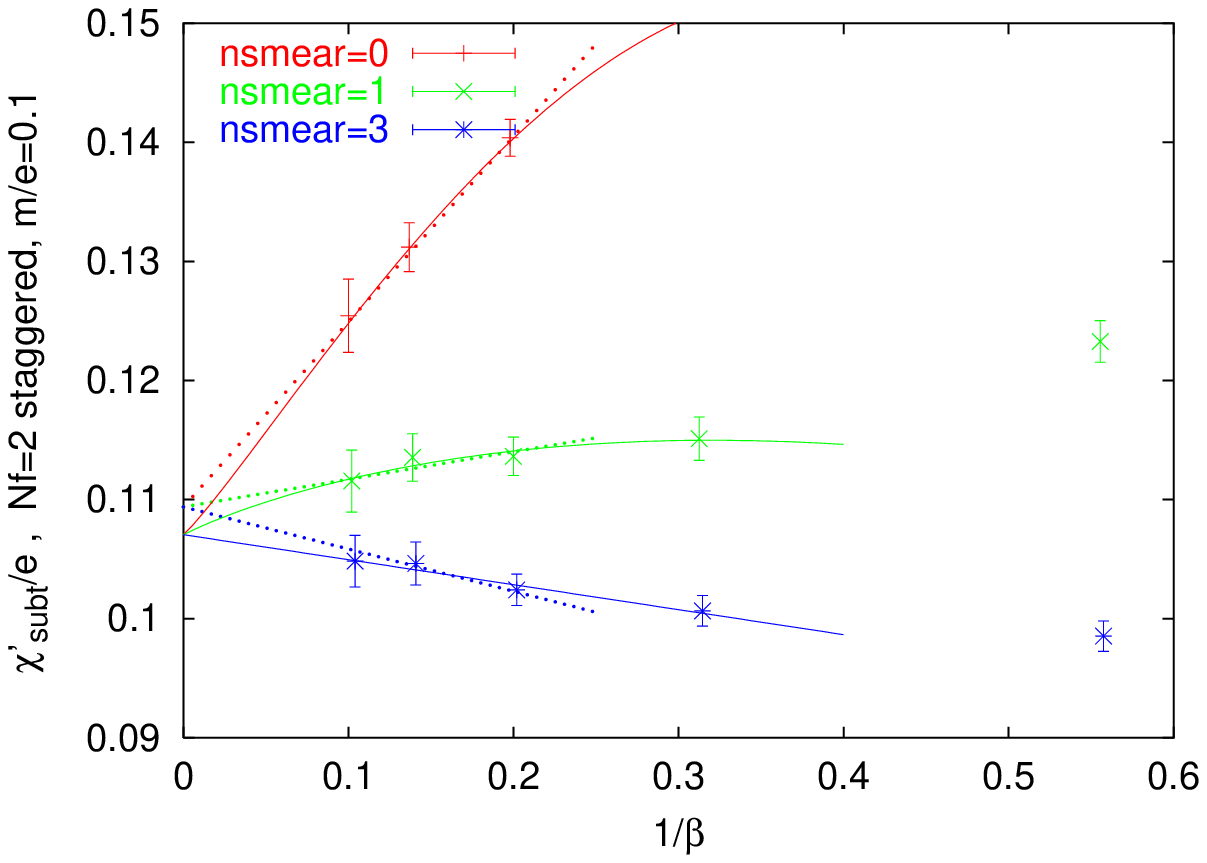,width=8.4cm}
\vspace{-8mm}
\caption{Continuum limit, via (\ref{ansatz_finimass}) without the log (full
lines), of $\ch_\mr{subt}'/e$ at fixed $m/e\!=\!0.1$ with overlap (left) and
staggered (right) quarks for $\Nf\!=\!0,1,2$ (top, center, bottom). Comparing to
the strictly linear extrapolation (dotted lines) gives a measure for the
systematic error.}
\label{fig:subt_pri_continuum}
\end{figure}

Having defined the ``primed'' subtracted condensates, we are in a position
to let, for any fixed $m/e\!>\!0$, the lattice spacing tend to zero
\bea
{\ch_\mr{cont}^{\prime\;\mr{ov}}(m/e)\ovr e}&=&
\lim_{1/\be\to0}\sqrt{\be}\ch_\mr{subt}^{\prime\;\mr{ov}}(m/e,1/\be)
\label{over_cont}
\\
{\ch_\mr{cont}^{\prime\;\mr{st}}(m/e)\ovr e}&=&
\lim_{1/\be\to0}\sqrt{\be}\ch_\mr{subt}^{\prime\;\mr{st}}(m/e,1/\be)
\label{stag_cont}
\eea
and Fig.~\ref{fig:subt_pri_allbeta} indicates that there is, at this stage, a
practical difference among the two formulations.
In the overlap case cut-off effects are mild, and this means that the continuum
limit can be taken, from the $\be$-values available, over the full mass range
shown.
With rooted staggered quarks, scaling violations get progressively worse for
small $m/e$.

Fig.~\ref{fig:subt_pri_continuum} indicates that, once the fermion mass is
large enough to be in the staggered scaling regime, both formulations agree in
the continuum, and the agreement seems not to be tied to a particular $\Nf$.
Comparing our ($\Nf\!=\!1$) continuum result
$\ch_\mr{cont}^{\prime\;\mr{ov}}(m/e\!=\!0.1)\!=\!0.142(4)$ to the values
0.1171 by Hosotani~\cite{Hosotani:1997pr} and 0.1277 by Adam~\cite{Adam:1998tw}
we find a disagreement at the $6.2\si$ or $3.6\si$ level, respectively.
This might indicate a finite-volume effect or a limited precision of the latter
calculations.
Since our boxlength is $\sim\!5$ times larger than the correlation length of
the lightest particle in the massless $\Nf\!=\!1$ theory, and in this regime
finite-volume effects are exponentially small~\cite{Sachs:en}, we consider the
first option with our data unlikely.

\begin{figure}[t!]
\begin{center}
\epsfig{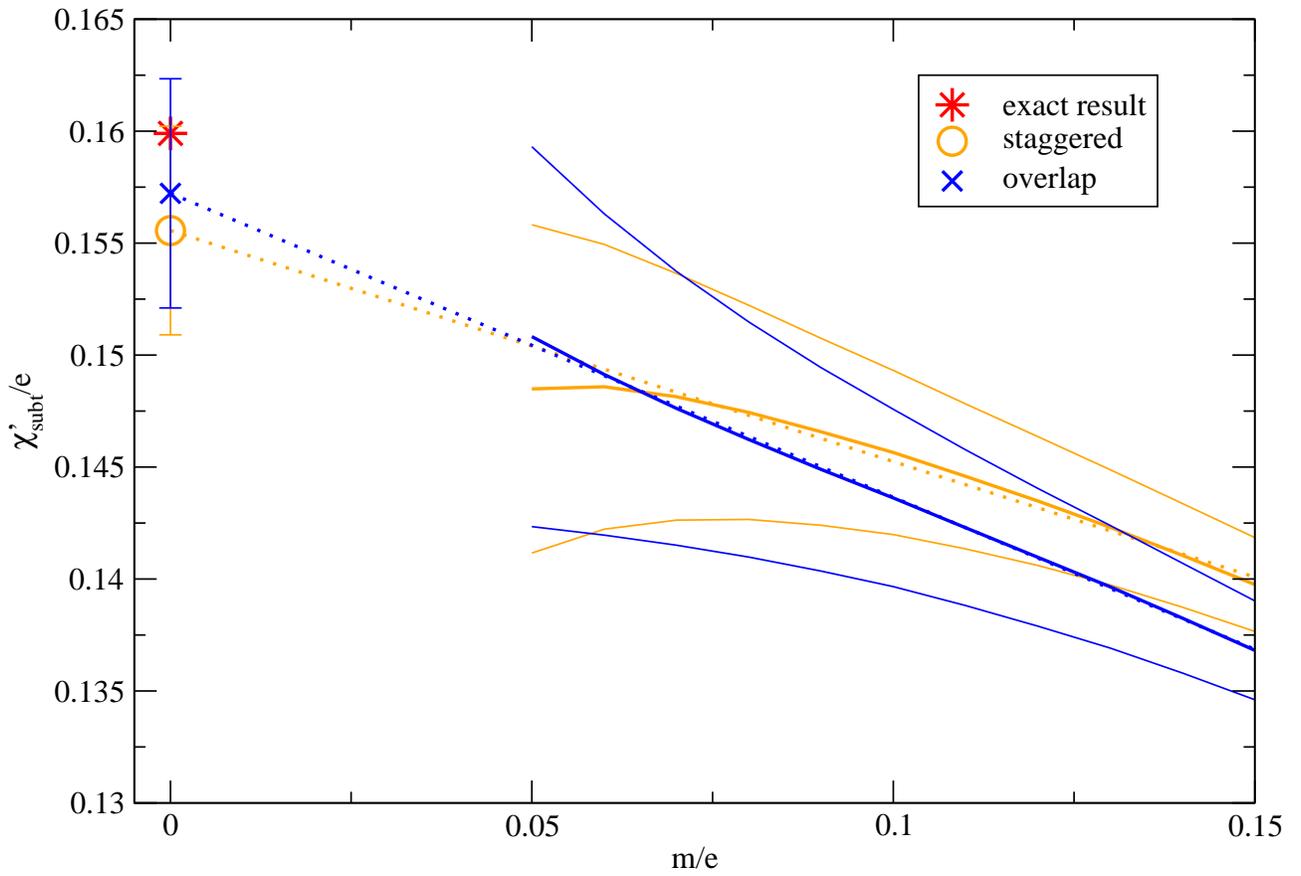}
\end{center}
\vspace{-8mm}
\caption{Chiral extrapolation of the continuum extrapolated primed condensates
(\ref{over_cont}, \ref{stag_cont}). In the staggered case the continuum limit
can only be taken for $m/e\!\geq\!0.05$ from our $\be$-values, and with overlap
quarks the same cut is applied for consistency. A linear extrapolation
$m\!\to\!0$ agrees with the analytical result (\ref{schwinger}) for either
discretization. The error-band is only statistical.}
\label{fig:cond_massextrapolation}
\end{figure}

As a last step, one my now take the chiral limit.
In the staggered case we find that after the $a\!\to\!0$ limit has been taken 
with the ansatz
\beq
{\chi_\mr{subt}\pri(m/e,1/\be,n)\ovr e}=A+B_n/\be+C\log(1/\be)+D_n/\be^{3/2}
\;.
\label{extrapolation}
\eeq
we have a continuum curve that covers the range $0.05\!\leq\!m/e\!\leq\!0.15$,
outside the ansatz (\ref{extrapolation}) yields an unacceptable $\ch^2$.
Using this segment and a linear ansatz for the extrapolation ${m\!\to\!0}$ we
find that the massless staggered $\Nf\!=\!1$ condensate agrees with the
analytical result (\ref{schwinger}).
In the overlap case the continuum curve would extend to $m/e\!=\!0$, but as a
consistency check we perform the same extrapolation, getting again a result
compatible with (\ref{schwinger}).
Both extrapolations are shown in Fig.~\ref{fig:cond_massextrapolation}.
Thus, for overlap fermions our data support the universal behavior
\beq
\lim_{a\to0}\lim_{m\to0}{\ch_\mr{subt}^{\prime\;\mr{ov}}(m/e,a^2)\ovr e}=
\lim_{m\to0}\lim_{a\to0}{\ch_\mr{subt}^{\prime\;\mr{ov}}(m/e,a^2)\ovr e}=
{\exp(\ga)\ovr2\pi^{3/2}}
\;.
\eeq
For staggered fermions, one the other hand, the suggested non-commutativity
phenomenon
\beq
\lim_{a\to0}\lim_{m\to0}{\ch_\mr{subt}^{\prime\;\mr{st}}(m/e,a^2)\ovr e}=0
\quad,\qquad
\lim_{m\to0}\lim_{a\to0}{\ch_\mr{subt}^{\prime\;\mr{st}}(m/e,a^2)\ovr e}=
{\exp(\ga)\ovr2\pi^{3/2}}
\label{stag_noncommutativity}
\eeq
means that rooted staggered fermions can be used to reproduce the Schwinger
result (\ref{schwinger}), but only if the right order of limits is chosen.


\section{Topological susceptibility}


Another interesting observable to study the effects of dynamical fermions is
the topological susceptibility which, in the context of this note, shall be
defined (in the continuum) through
\beq
\ch_\mr{top}=\lim_{V\to\infty}
{\<\det(D\!+\!m)^\Nf\;q^2\>\ovr V\,\<\det(D\!+\!m)^\Nf\>}
\;.
\label{defsusc}
\eeq
The main difference to the scalar condensate is that the topological
susceptibility depends only on the \emph{sea\/}-quarks, thus offering a
potentially cleaner view at the effects of square-rooting the staggered
determinant to get $\Nf\!=\!1$.

\begin{figure}[b!]
\begin{center}
\epsfig{file=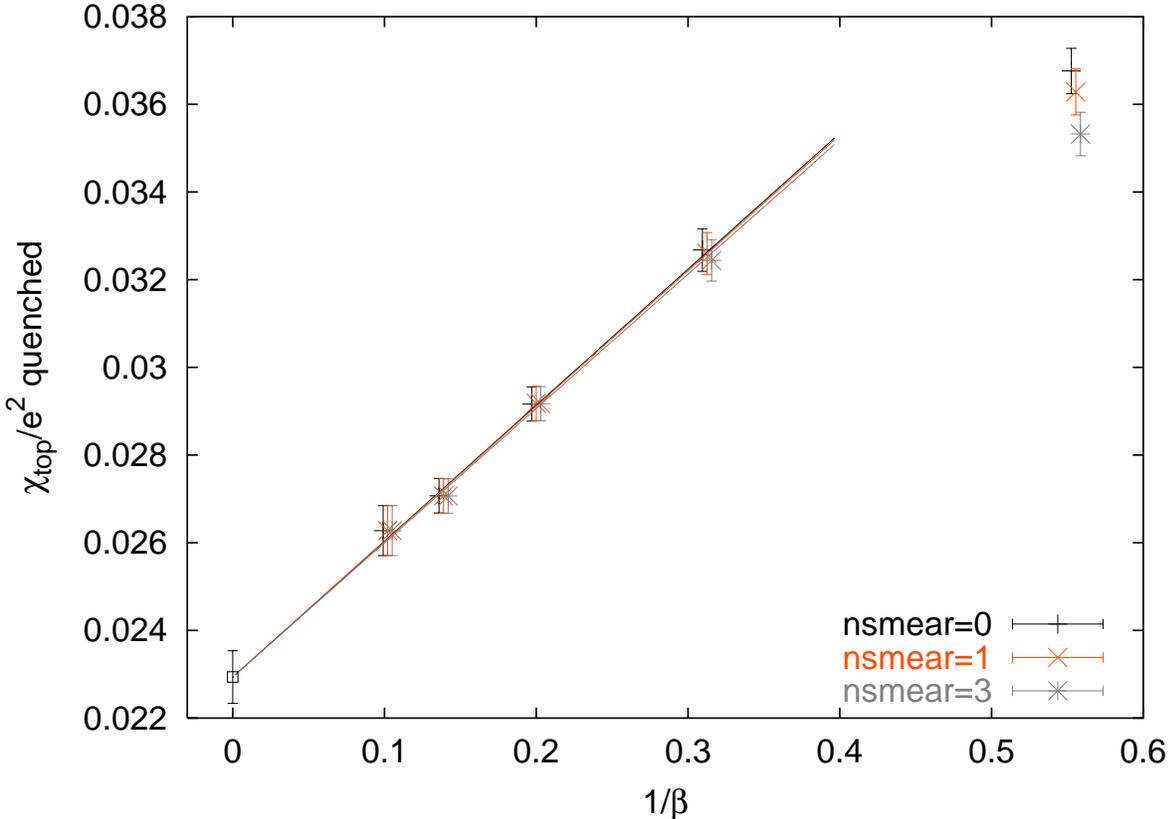,width=16cm}
\end{center}
\vspace{-8mm}
\caption{The zero temperature quenched topological susceptibility
$\ch_\mr{top}^{\Nf=0}/e^2$ versus $(ae)^2$. The symbols for
$\mr{nsmear}\!=\!0,3$ are slightly offset for better visibility. Note the steep
slope -- in spite of being in the scaling regime for $\be\!\geq\!3.2$,
cut-off effects can be large. At $\be\!=\!9.8$ the charge $q$ is independent of
the smearing level for all our configurations.}
\label{fig:susc_quenched}
\end{figure}

\begin{figure}
\begin{center}
\epsfig{file=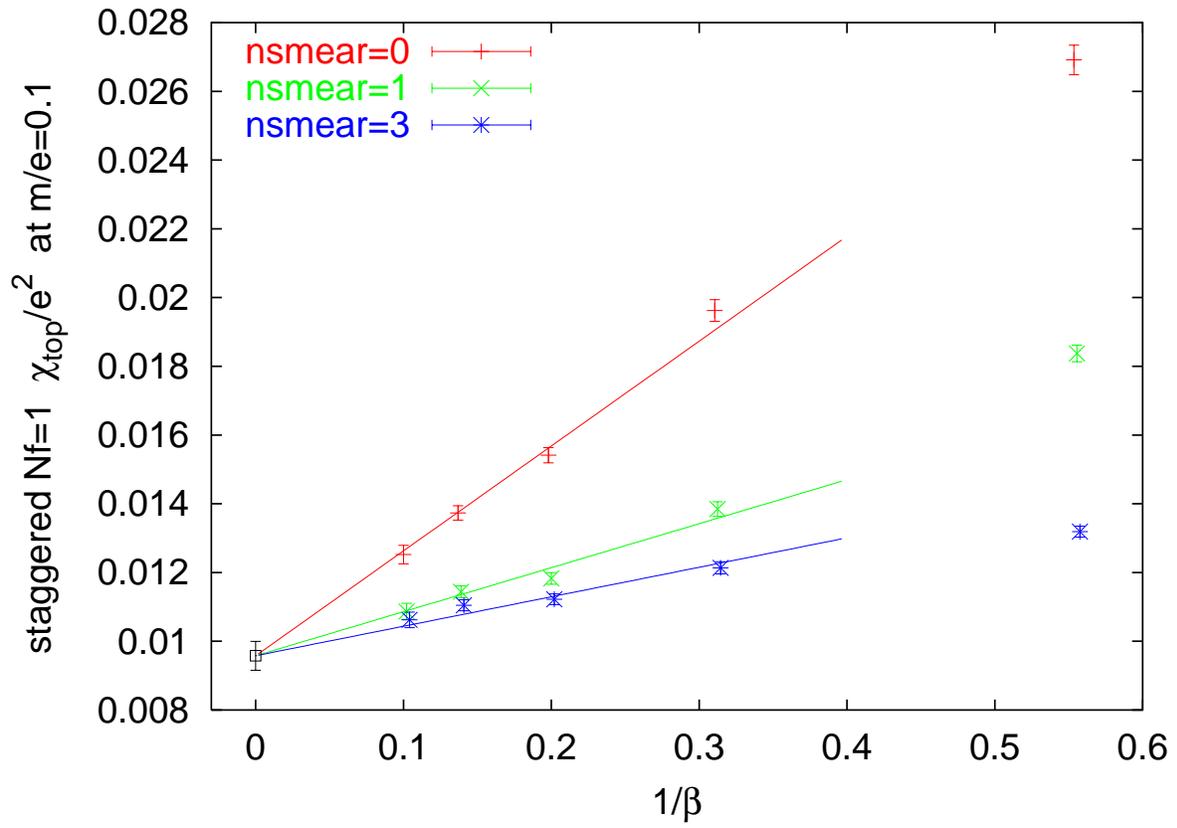,width=16cm}
\epsfig{file=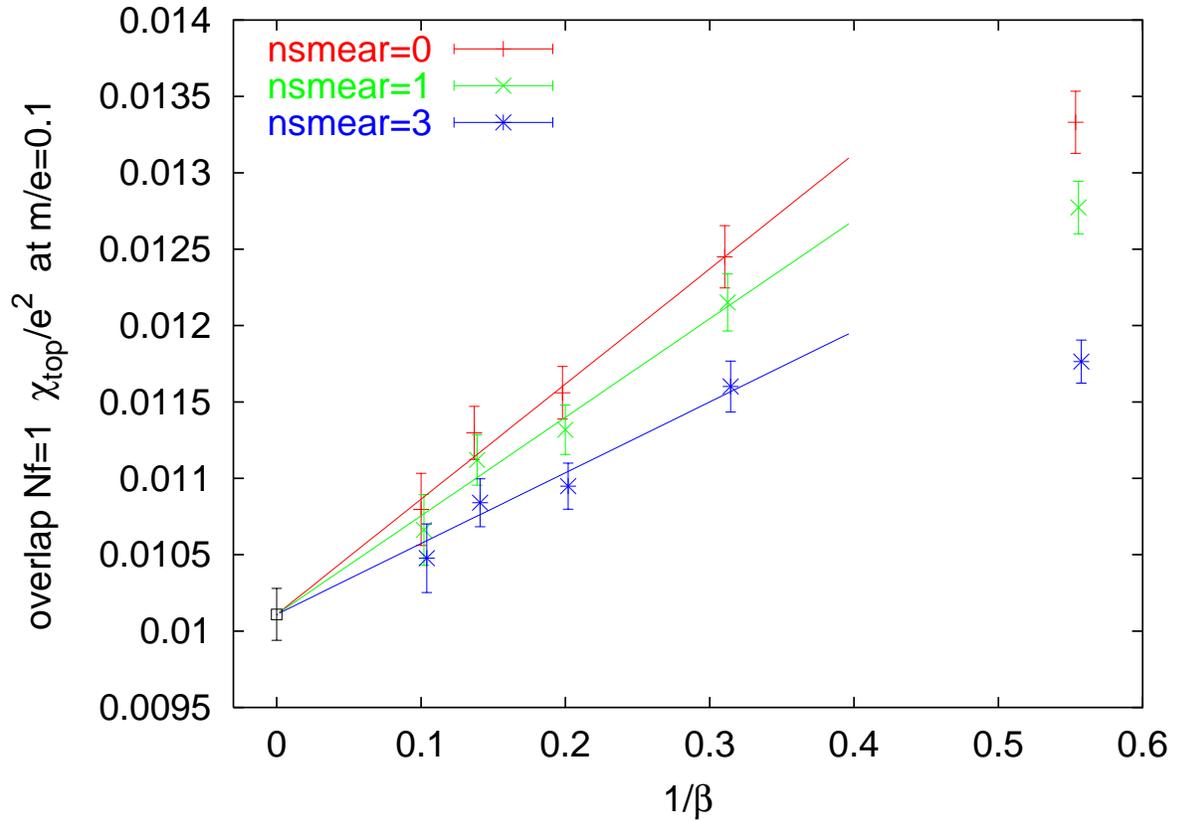,width=16cm}
\end{center}
\vspace{-8mm}
\caption{At the fixed quark mass $m/e\!=\!0.1$ a correlated fit to all
staggered $\Nf\!=\!1$ data (top) suggest a continuum limit 0.00958(42) which is
consistent with 0.01011(17) from all overlap data (bottom).
In the staggered case, the first smearing renders the extrapolation much
flatter, while the overlap has a flat behavior with any level of filtering
(note the difference in scale).}
\label{fig:susc_continuum_nf1}
\end{figure}

\begin{figure}
\begin{center}
\epsfig{file=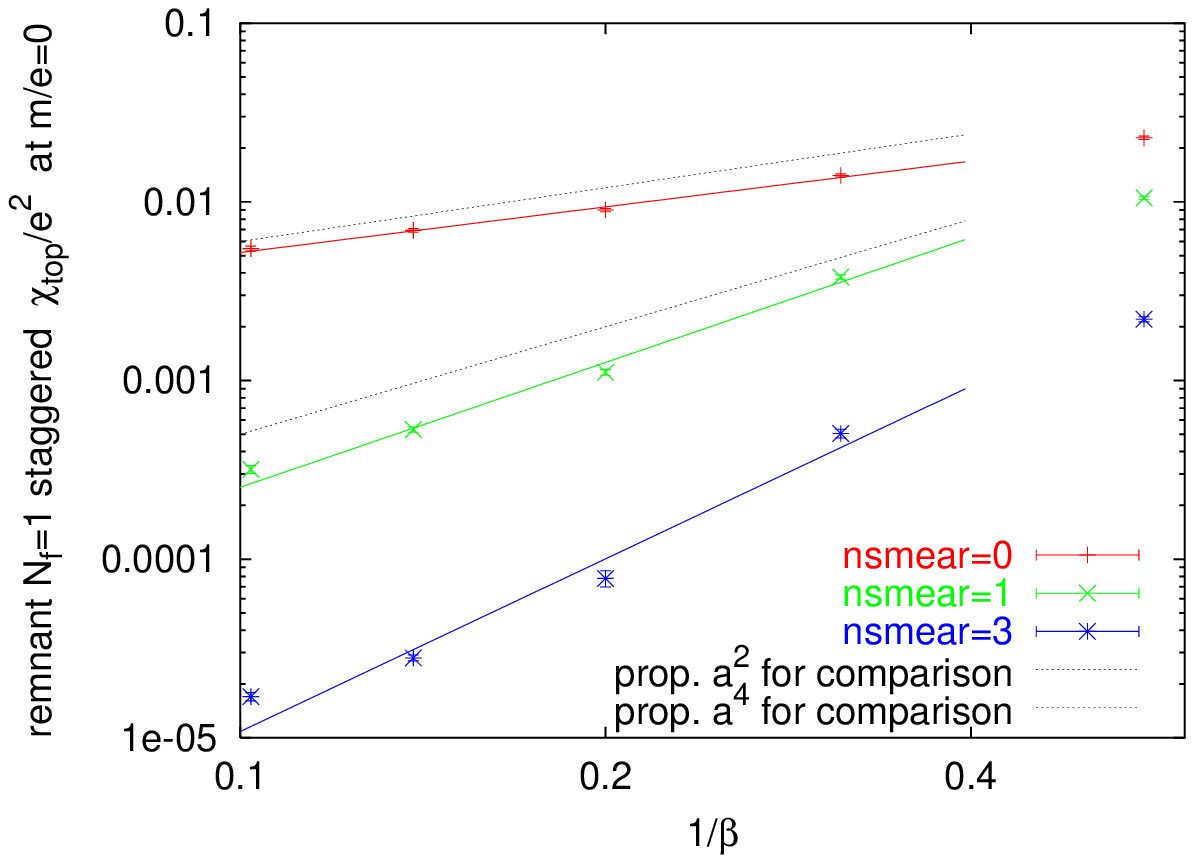,width=8.4cm}
\epsfig{file=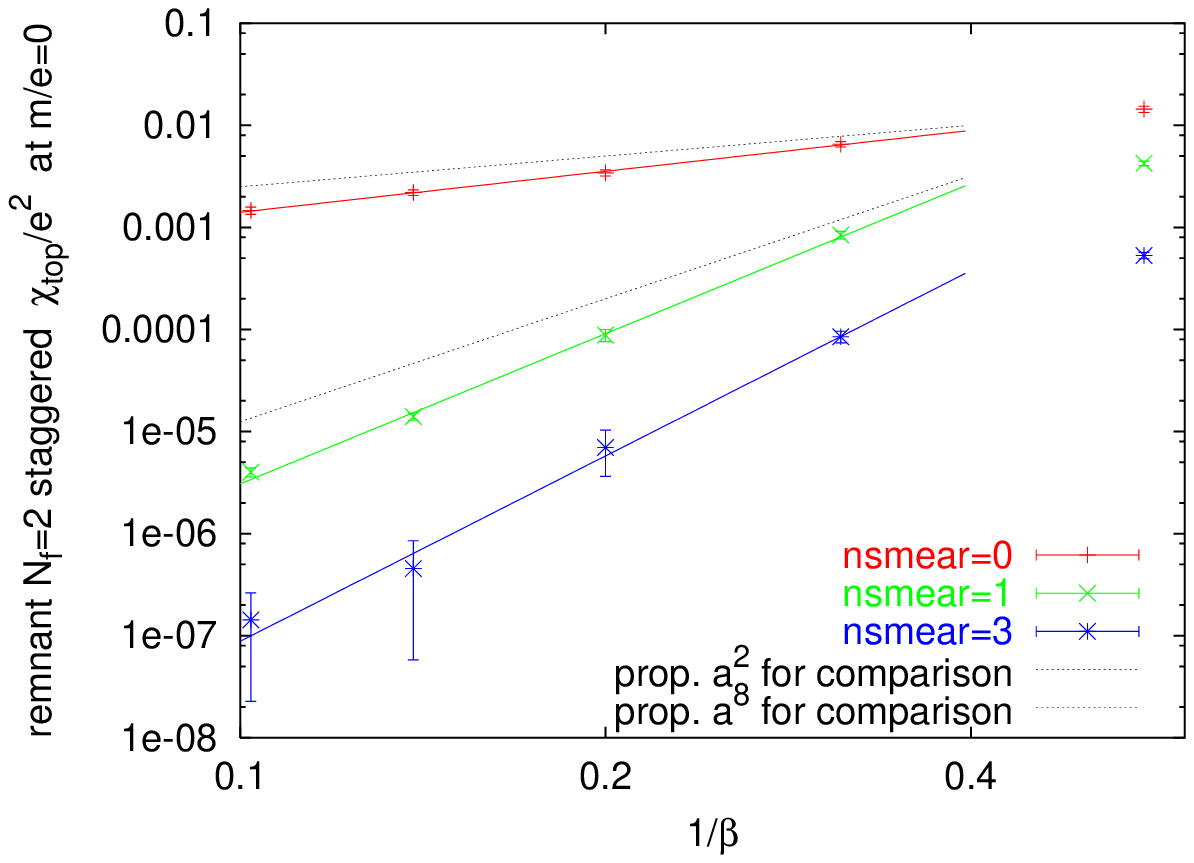,width=8.4cm}
\end{center}
\vspace{-8mm}
\caption{The remnant $\ch_\mr{top}^\mr{st}/e^2$ at $m/e\!=\!0$ versus
$1/\be$, in log-log scale. This staggered artefact seems to disappear with a
power-law in $a$, for $\Nf\!=\!1$ (left) and $\Nf\!=\!2$ (right), but the
power is 2 only for smearing level 0, while it is $4...8$ for higher filtering
levels (over our range of $\be$-values).}
\label{fig:susc_remnant}
\end{figure}

For staggered quarks, the definition (\ref{defsusc}), taken in fixed volume,
reduces to
\beq
{\ch_\mr{top}^\mr{st}\ovr e^2}=
{\be\ovr L^2}\,{\<\det(D^\mr{st}_m)^{\Nf/2}\;q^2\>\ovr
\<\det(D^\mr{st}_m)^{\Nf/2}\>}
\;,
\label{suscstag}
\eeq
while for overlap quarks, the implementation is
\beq
{\ch_\mr{top}^\mr{ov}\ovr e^2}=
{\be\ovr L^2}\,{\<\det(D^\mr{ov}_m)^\Nf\;q^2\>\ovr
\<\det(D^\mr{ov}_m)^\Nf\>}
\;,
\label{suscover}
\eeq
where $\det(D^\mr{st}_m)$ and $\det(D^\mr{ov}_m)$ are defined in
(\ref{condstageig}, \ref{condovereig}).
The sum is over the full spectrum of the massless operator, and we apply the
overlap definition (\ref{defindex}) of the charge $q$ in either case.

Let us begin with a check that a reasonable number of our zero temperature
$\be$-values is in the scaling regime before reweighting.
Fig.~\ref{fig:susc_quenched} contains our quenched topological susceptibility
data, and $\be\!\geq\!3.2$ seems sufficient to be in the regime with
only $O(a^2)$ effects.

The first test is whether overlap and rooted staggered quarks yield the same
result in the continuum, if the topological susceptibility
$\ch_\mr{top}^{\Nf=1}(m/e)/e^2$ is taken at a fixed non-vanishing quark mass.
Fig.~\ref{fig:susc_continuum_nf1} presents the outcome for $m/e\!=\!0.1$.
In either case the three smearing levels seem to have a common continuum
limit and this is why we adopt a correlated linear fit.
The staggered continuum value 0.00958(42) is consistent with 0.01011(18) from
all overlap data.
This is of course not a proof but good numerical evidence that rooted staggered
quarks yield the correct continuum limit for the topological susceptibility at
finite quark mass.

The second test concerns the chiral limit, where $\ch_\mr{top}^\mr{ov}(0)\!=\!0$
(as in the continuum~\cite{Leutwyler:1992yt}), while
$\ch_\mr{top}^\mr{st}(0)\!>\!0$ is a pure discretization effect.
Fig.~\ref{fig:susc_remnant} shows the remnant staggered susceptibility
at $m/e\!=\!0$ versus $(ae)^2$, in a log-log representation.
With unfiltered staggered sea-quarks it seems to disappear in proportion to
$a^2$, regardless of $\Nf$.
For the filtered variety we find a slope $\sim\!2$ (i.e.\ dominating $O(a^4)$
cut-off effects) over the range of accessible couplings for $\Nf\!=\!1$ and
even $\sim\!4$ (i.e.\ dominating $O(a^8)$ effects) for $\Nf\!=\!2$.
Still, it is conceivable that this slope eventually flattens out and the
\emph{asymptotic\/} cut-off effects might be $O(a^2)$ with any level of
filtering.
This would mean that smearing renders the coefficient in front of the $O(a^2)$
term so small that the ``subleading'' $O(a^4)$ or $O(a^8)$ terms would
numerically dominate over a substantial range of couplings.
In any case, the important news is that the non-commutativity phenomenon in the
chiral condensate is not replicated here -- for the topological susceptibility
the staggered answer is correct even if the chiral limit is taken before the
continuum limit.


\section{Partition function and Leutwyler-Smilga sum rules}


In the $\ep$-regime of QCD~\cite{Leutwyler:1992yt} ($m\Sigma V\!\ll\!1$, but
still with a large box-size, i.e.\ $L\!\gg\!1/(2\Fpi)\!\simeq\!1\fm$, c.f.\ the
discussion in~\cite{Colangelo:2003hf}) the log of the partition function is
known analytically~\cite{Leutwyler:1992yt}
\beq
Z_\th(m)=e^{m\Sigma V\cos(\th)}
\label{LS1}
\eeq
with sub-leading corrections of order $(m\Sigma V)^2$.
The partition function in a sector of fixed topological charge $q$ follows by
taking the Fourier transform~\cite{Leutwyler:1992yt}
\beq
Z_q(m)=\int d\th\;e^{\ri q\th}Z_\th(m)=I_{|q|}(m\Sigma V)
\label{LS2}
\eeq
where $I_{|q|}$ is the modified Bessel function.

\begin{figure}
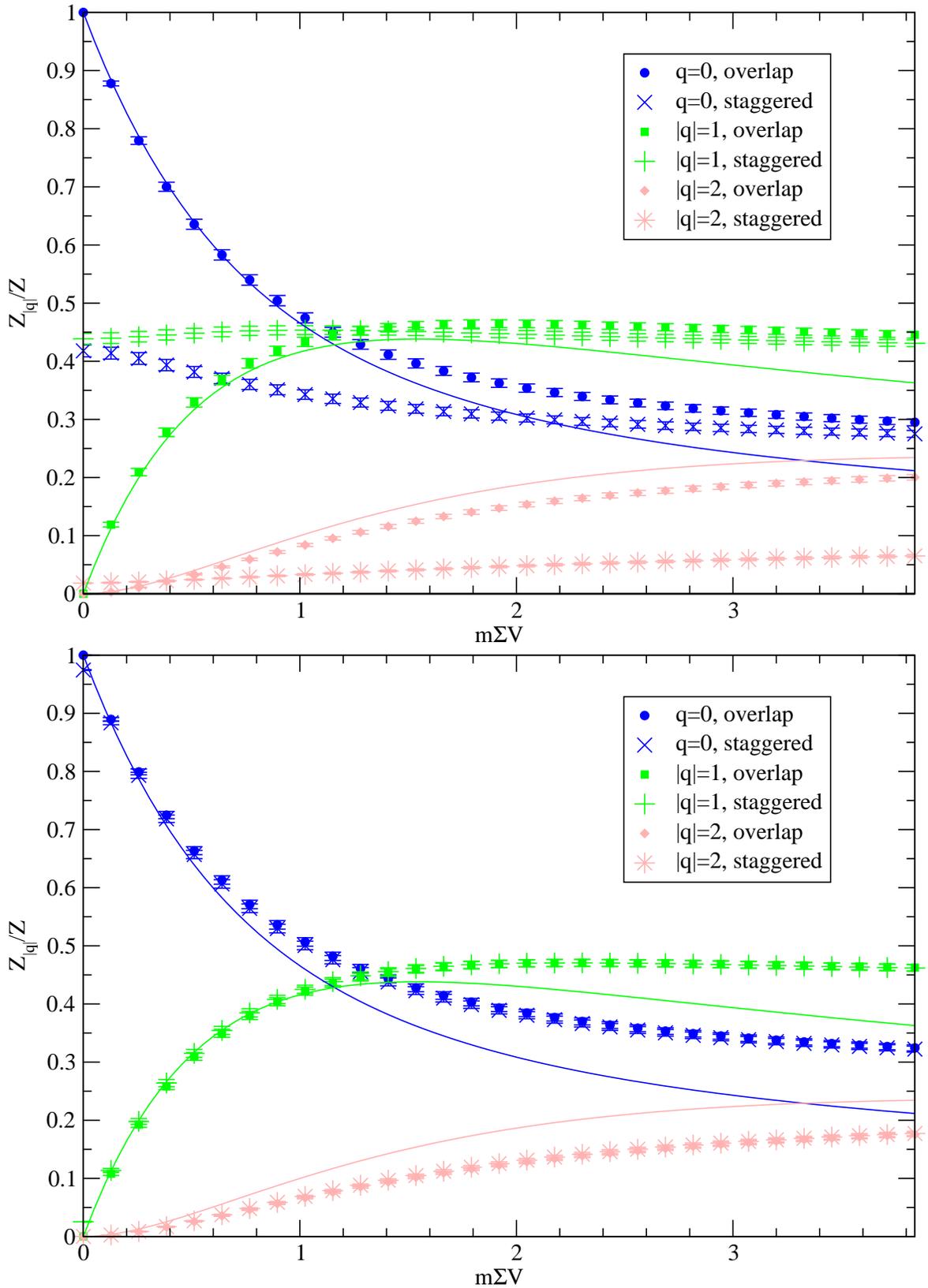

\begin{center}
\epsfig{file=smrw_v2.figs/pfo_0_3.2.eps,width=15.5cm}
\epsfig{file=smrw_v2.figs/pfo_1_9.8.eps,width=15.5cm}
\end{center}
\vspace{-8mm}
\caption{The $\Nf\!=\!1$ partition function $Z_{|q|}$ (normalized by the sum
over all $q$) at $\be\!=\!3.2$ without filtering (top) and at $\be\!=\!9.8$
with 1 smearing step (bottom) versus $x\!=\!m\Sigma V$. For sufficiently small
mass the overlap version (filled symbols) is always a decreasing function of
$|q|$, as in the continuum. The rooted staggered version (crosses) joins, for
large mass, the associated overlap $Z_{|q|}/Z_{\th=0}$. The full lines
represent the parameter-free prediction (\ref{LS1}, \ref{LS2}).}
\label{fig:partfunc}
\end{figure}

\begin{figure}
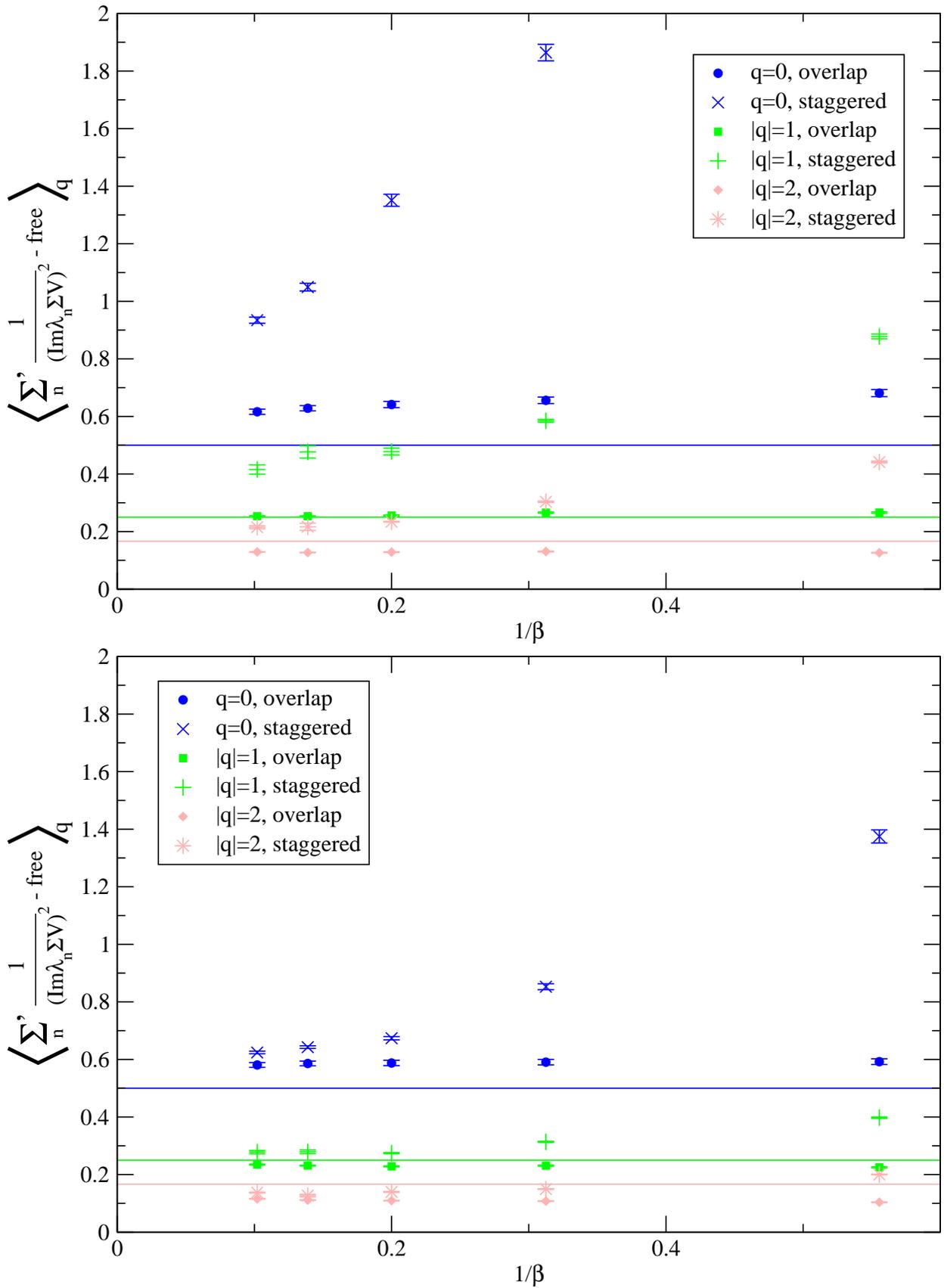

\begin{center}
\epsfig{file=smrw_v2.figs/ls_0_sub.eps,width=16.5cm}
\epsfig{file=smrw_v2.figs/ls_1_sub.eps,width=16.5cm}
\end{center}
\vspace{-8mm}
\caption{The Leutwyler-Smilga sum rule (\ref{LS4}) for $\Nf\!=\!1$ active
quarks versus $(ae)^2$, at smearing level 0 (top) and 1 (bottom).
Note the dramatic difference in the scaling behavior between overlap and rooted
staggered fermions. The deviation, in the continuum, from the Leutwyler-Smilga
prediction (full lines) is likely a finite-volume effect.}
\label{fig:sumrule}
\end{figure}

By differentiating with respect to the quark mass and setting the latter to
zero, Leutwyler and Smilga obtained the first sum rule for the inverse
eigenvalues of the massless Dirac operator
\beq
\left\<
\sum_n\nolimits^\prime{1\ovr(\mr{imag}(\la_n)\Sigma V)^2}
\right\>_{\!\!q}^{\!\!\prime}=
{1\ovr2(|q|\!+\!\Nf)}
\label{LS3}
\eeq
where the primes indicate that in a sector with topological charge $q\!\neq\!0$
the sum is over the full spectrum without the $|q|$ zero modes and the
importance sampling is to be carried out with the ``primed determinant'', i.e.\
without the zero modes.
In 4D the l.h.s.\ is quadratically divergent, since the eigenmode distribution
is asymptotically proportional to $\la^3$.
In 2D the divergence is logarithmic, since the eigenmode distribution is
asymptotically linear in $\la$ [cf.\ (\ref{cond_continuum})].
One way out is to subtract the free case in the same volume.
An other one is to consider the difference between two topological
sectors, since the UV-behavior is independent of the charge.

We start with the \emph{conjecture\/} that eqn.\ (\ref{LS1}) and thus
(\ref{LS2}, \ref{LS3}) hold in the 1-flavor Schwinger model, too, albeit with
re-interpreting $\Sigma$ as the analytically known 1-flavor condensate
(\ref{schwinger}).

In Fig.~\ref{fig:partfunc} we show the $\Nf\!=\!1$ partition function on a
coarse and a fine lattice, without and with one APE step,
respectively.
In the first case, one sees a significant difference between the overlap and
the rooted staggered answer for small quark masses.
For larger $\be$ and with just one smearing step this difference disappears
and we did not find any indication for a disagreement in the continuum limit
at any $m/e$.
For $x\!\ll\!1$ the continuum extrapolated data agree well with
$(2\!-\!\de_{q,0})I_{|q|}(x)/\exp(x)$ which, with our choice
for $\Sigma$, is a parameter-free prediction.
We like to point out that this figure is reminiscent of the situation in QCD,
see Fig.~3 of Ref.~\cite{Ogawa:2004ru}.

To verify (\ref{LS3}) we subtract the free case, i.e.\ we check whether
\beq
\left\<
\sum_n\nolimits^\prime{1\ovr(\mr{imag}(\la_n)\,\Sigma V)^2}-
\sum_n\nolimits^\prime{1\ovr(\mr{imag}(\la_n^\mr{free})\,\Sigma V)^2}
\right\>_{\!\!q}^{\!\!\prime}=
{1\ovr2(|q|\!+\!\Nf)}
\label{LS4}
\eeq
with $\la_n$ denoting the (purely imaginary) $n$-th staggered or chirally
rotated overlap eigenvalue (i.e.\ the $\hat\la$ of (\ref{condovereig2})), and
in the staggered case an additional factor $1/2$ on the l.h.s.\ is needed (or
$1/4$ in 4D).
The prime on the sum indicates that on topologically nontrivial configurations
the $|q|$ true overlap zero modes or the $|q|$ ``would be'' zero modes on either
side of the staggered spectrum are excluded.
Since also the free overlap and staggered Dirac operators have 2 and 4
(non-topological) zero modes, respectively, the $\la_n^\mr{free}$ denote the
free eigenvalues without these zeros.
Thus the sum over $n$ does not include the same number of terms in the free and
interacting case.
The prime above the expectation value indicates that in a $q\!\neq\!0$ sector
the $|q|$ zeros are removed from the determinant.
Fig.~\ref{fig:sumrule} contains our results for the sum rule (\ref{LS4}), where
our choice for the interpretation of $\Sigma$ leads again to a parameter-free
prediction.
The deviation from the Leutwyler-Smilga value is likely a finite-volume effect.
In any case, there is no sign of a difference between the rooted staggered
and the overlap answer in the continuum.


\section{Heavy quark potential}


The heavy-quark (HQ) potential is an interesting observable, since it is easy
to measure in the Schwinger model and the effect of dynamical fermions is
clearly visible.
We define the HQ potential $V(r)$ in a finite volume $L\times T$ via the
Polyakov loop correlator
\beq
e^{-V(r)T}=\<P(0)P\dag(r)\>
\;.
\label{def_polyakov}
\eeq

In the quenched case, and with $T\!\to\!\infty$, it has a simple saw tooth
shape, i.e.\ it starts at zero, raises linearly for $0\!\leq\!r\!\leq\!L/2$,
and then it decreases linearly, until it reaches $0$ at $r\!=\!L$ again.
This follows from the fact that the quenched Wilson loop in infinite volume
satisfies an exact area law $W_{\Nf=0}(r,t)=(I_1(\be)/I_0(\be))^{rt}$ [in
lattice units], thus the force in physical units is
\beq
{F_{\Nf=0, T=\infty}(r)\ovr e^2}=-\be\log\Big({I_1(\be)\ovr I_0(\be)}\Big)
\;.
\label{HQforce_nf0}
\eeq

In the $\Nf\!=\!1$ case an exact solution on the torus is known for $m\!=\!0$
\cite{Sachs:en}, but --~as far as we know~-- not for massive fermions.
However, we expect the screening behavior found in the massless case to also
apply, on a qualitative level, for a small enough fermion mass.

\begin{figure}
\begin{center}
\epsfig{file=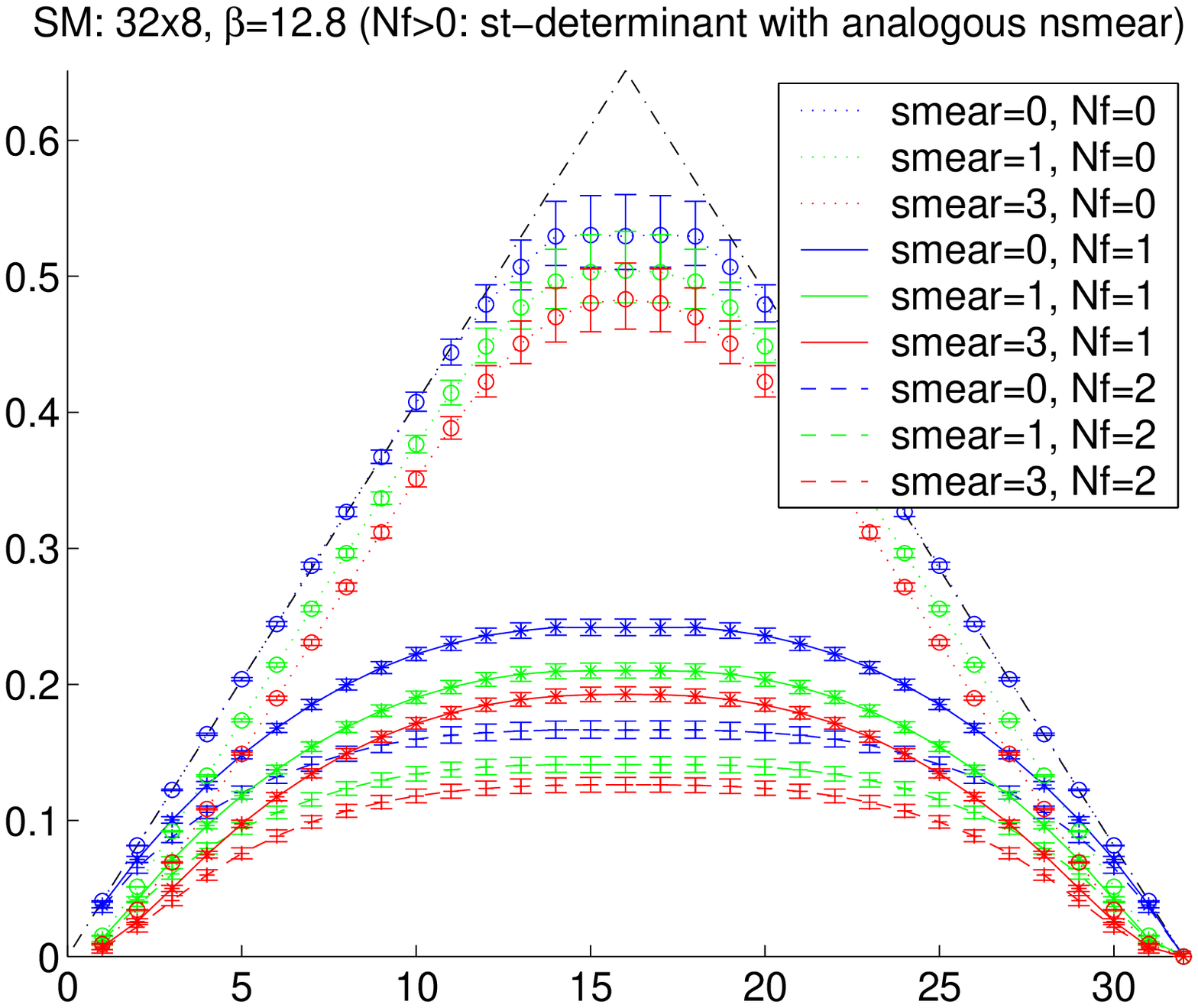,width=13.7cm}
\epsfig{file=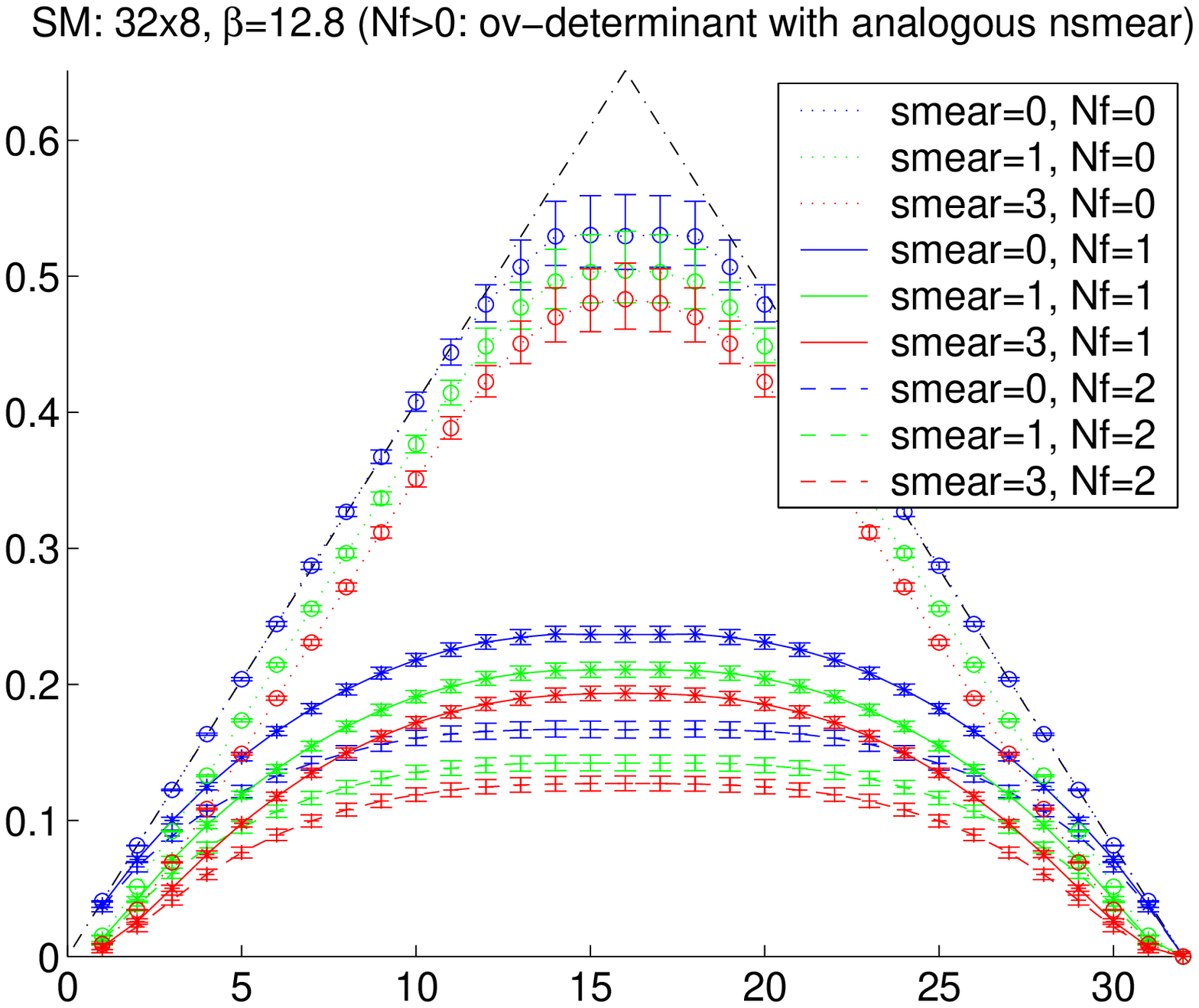,width=13.7cm}
\end{center}
\vspace{-8mm}
\caption{The quenched zero temperature HQ-potential is a saw tooth. After
reweighting with staggered (top, rooted for $\Nf\!=\!1$) or overlap (bottom)
fermions ($m/e\!=\!0.1$), screening is seen. Throughout, the same smearing is
applied in the HQ line and in the fermion operator. Except near the endpoints,
the improved HQ action leads to a mere downwards shift of the potential.}
\label{fig:HQ_12.8}
\end{figure}

\begin{figure}
\begin{center}
\epsfig{file=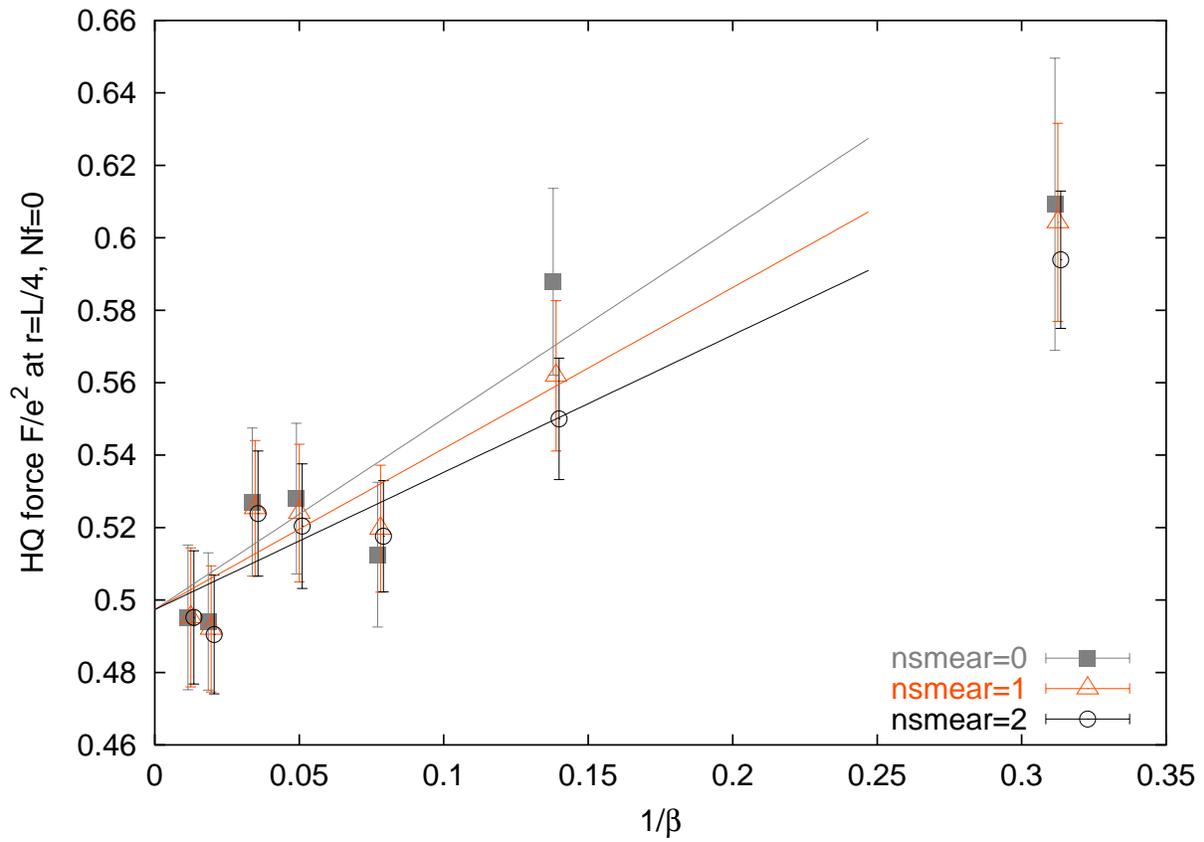,width=16.2cm}
\epsfig{file=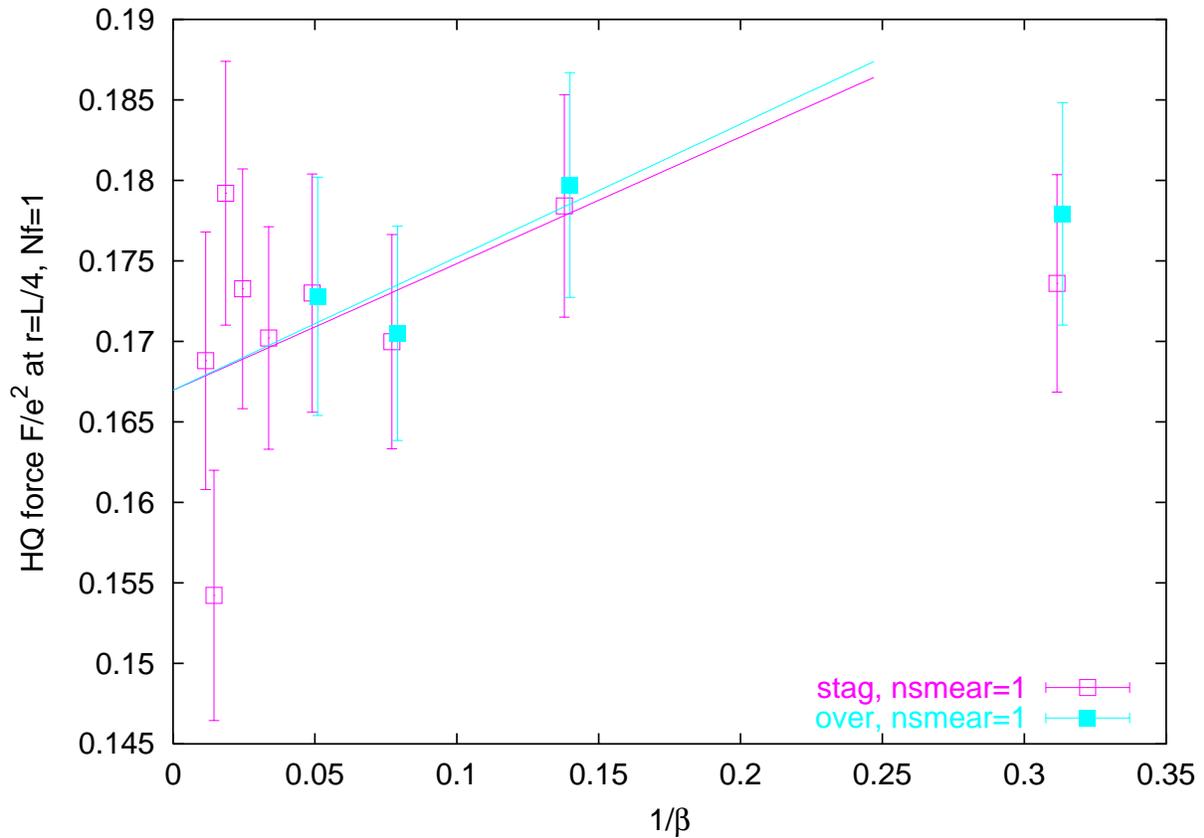,width=16.2cm}
\end{center}
\vspace{-8mm}
\caption{Top: With our $\be$-values, the quenched heavy-quark force $F/e^2$ at
$r\!=\!L/4$ is in the scaling regime. For large enough $\be$, the three
HQ actions agree in their central values and errors. Bottom: The $\Nf\!=\!1$ HQ
force $F/e^2$ at $r\!=\!L/4$ with overlap or rooted staggered fermions
(filtering level 1 applied in both and the HQ action) seems to have a universal
continuum limit.}
\label{fig:force_nf0nf1}
\end{figure}

\begin{figure}[t]
\begin{center}
\epsfig{file=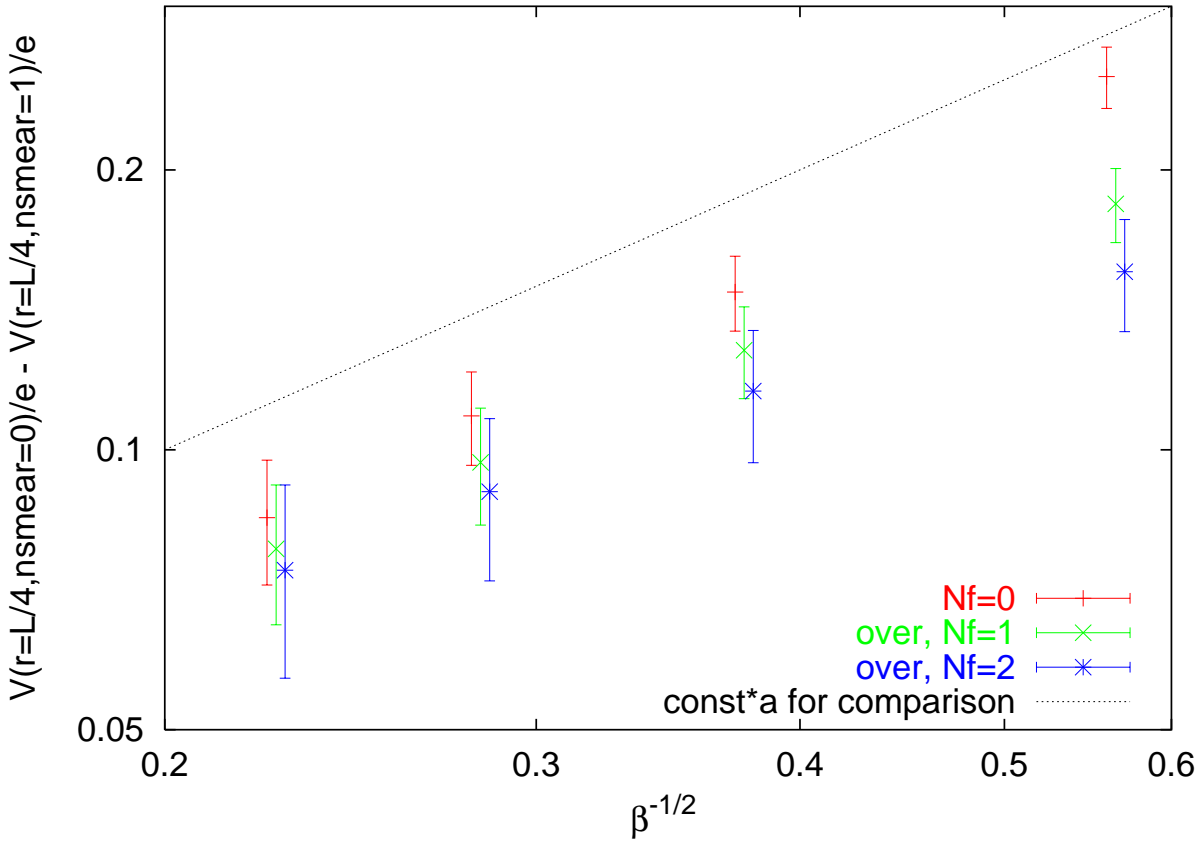,width=16.5cm}
\end{center}
\vspace{-8mm}
\caption{The difference between the HQ potentials with smearing-level 0 and 1
(at $r\!=\!L/4$) versus $ae$, in log-log scale. It
disappears in proportion to $a$ -- the offset line has slope 1.}
\label{fig:HQ_shift}
\end{figure}

In Fig.~\ref{fig:HQ_12.8} we show an example of our HQ potentials
($\be\!=\!12.8$, lattice units).
In this regime the quenched prediction (\ref{HQforce_nf0}) works very well,
except that the finite temperature causes a flattening of the peak region.
If the Polyakov loops are built from smeared links instead of the original
ones (``improved HQ action'', see~\cite{Lepage:1991ui,HQselfenergy4D}),
the potential is merely shifted downwards, except near $r=0,L$ where some
curvature is introduced.
In other words, the force is unaffected by such a modification of the HQ
action, provided it is measured at a distance larger than the number of
smearing steps applied.
Furthermore, the reduced HQ selfenergy is supposed to lead to a smaller error
bar (originally~\cite{Lepage:1991ui}, see ~\cite{HQselfenergy4D} for details),
but here it seems unaffected -- we will come back to this point.
After reweighting to $\Nf\!=\!1$ or $\Nf\!=\!2$, screening is seen (as
expected), yet there is no visible difference between staggered (top) or
overlap (bottom) dynamical flavors.
Here we show only the ``diagonal'' data, where the same smearing level is
applied in the HQ action as in the fermion operators, but the ``offdiagonal''
data are similar.

For a detailed analysis we shall concentrate on the force at a given
(large enough) distance, and we choose to present a scaling analysis for
$F_{\Nf=1}(r\!=\!L/4)/e^2$.
Before doing so, it is worthwhile to check that our $\be$-values are such that
the quenched counterpart, $F_{\Nf=0}(r\!=\!L/4)/e^2$, is in the scaling regime,
and the top of Fig.~\ref{fig:force_nf0nf1} shows that this is indeed the case
-- regardless of the HQ action (smearing level of the Polyakov loop links)
used.
When employing overlap or rooted staggered fermions to have one active flavor
with $m/e\!=\!0.1$, nothing fundamental changes, as shown in the bottom of
Fig.~\ref{fig:force_nf0nf1}.
In fact, at the 4 couplings where we have data with either discretization,
the results are fully compatible already at finite $\be$.
Thus, in the case of the HQ force a rooted staggered field seems to yield the
same continuum limit as a single overlap flavor.

Let us finally discuss an interesting aside.
It has been argued (originally~\cite{Lepage:1991ui}, see ~\cite{HQselfenergy4D}
for details) that the selfenergy $\de\!m$ of a quark in the static
approximation is directly related to the noise in an observable with such a
quark.
The important point is that in 4D the selfenergy in physical units diverges
linearly in the inverse lattice spacing~\cite{Eichten:1989zv}
\beq
\de\!m\!=\!\mr{const}/a
\qquad[\mr{4D}]\;.
\eeq
If changing the discretization amounts to a replacement
$\mr{const}^{'}\!\to\!\mr{const}^{''}$, then smearing becomes
\emph{more important\/} on \emph{fine\/} lattices.
What we wish to point out is that the situation in 2D is just opposite.
Here, the selfenergy vanishes in proportion to the lattice spacing
\beq
\de\!m\!=\!\Lambda_\mr{HQ}^2\,a
\qquad[\mr{2D}]\;.
\eeq
As a consequence, the usefulness of smearing decreases in 2D towards the
continuum.
This is clearly seen in the top of Fig.~\ref{fig:force_nf0nf1}.
In the rightmost point ($\be\!=\!3.2$) the smeared HQ action leads to a smaller
error bar, while this effect quickly disappears towards the left.
This is matched by the behavior of the shift of the HQ potential [in physical
units], brought by a single smearing step, as shown in Fig.~\ref{fig:HQ_shift}.
Such a pattern is expected, if the first smearing step amounts to a replacement
$\Lambda_\mr{HQ}^{'}\!\to\!\Lambda_\mr{HQ}^{''}$, and it is reassuring to see
that this rule holds for any $\Nf$.


\section{Determinant ratios}


The last topic that we wish to discuss is how well the rooted staggered
determinant manages to approximate the (one flavor) overlap determinant.
It has been shown --~both in 2D~\cite{DuHo} and in 4D~\cite{DuHoWe}~-- that
the \emph{low energy\/} eigenvalues eventually coincide (apart from an overall
rescaling factor and the $2^{d/2}$-fold degeneracy), in the limit of fine
lattice spacings.
However, in the UV region the two spectra are very different, and, if the
resolution is made finer in a fixed physical volume, the total number of modes
grows in proportion to $1/a^d$.
Therefore, it is not clear whether the full one-flavor determinants would
eventually coincide.
In fact, $\det(D^\mr{ov})$ and the $2^{d/2}$-fold root of $\det(D^\mr{st})$
are never equal (typically, they differ by many orders of magnitude), but the
right question to ask is whether the two formulations give the same answer
--~modulo cut-off effects~-- for the determinant ratio on two arbitrary
configurations, i.e.\ whether
\beq
{\det(D^\mr{ov}(U))\ovr\det(D^\mr{ov}(U'))}=
\sqrt[2^{d/2}]{\det(D^\mr{st}(U))\ovr\det(D^\mr{st}(U'))}
\;\Big(1+O(a^2)\Big)
\;.
\label{detratio}
\eeq

\begin{figure}[t]
\begin{center}
\epsfig{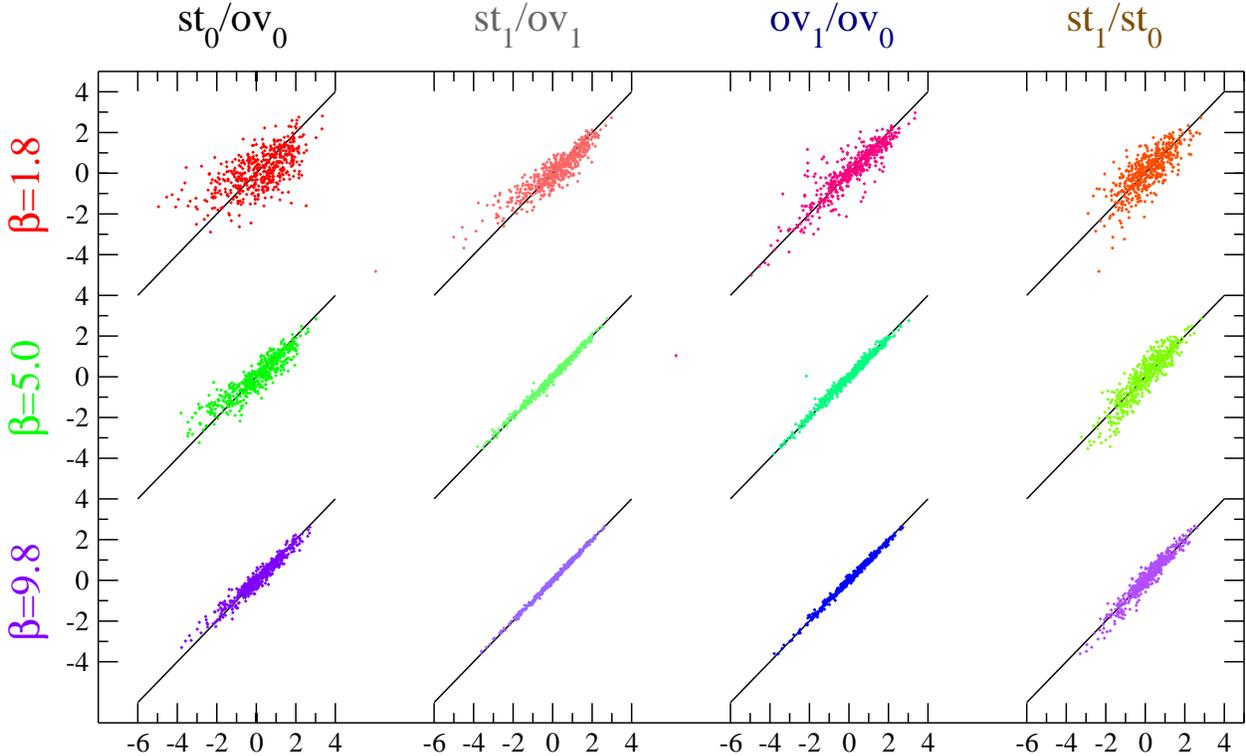}
\end{center}
\vspace{-6mm}
\caption{Log-log plot of the rooted staggered versus the overlap determinant,
at fixed physical mass $m/e\!=\!0.1$ and $\be\!=\!1.8, 5.0, 9.8$ (from top to
bottom), without (leftmost column) and after one step of filtering (second
column). In addition, we plot smearing level~1 versus level~0 for the overlap
(third) and the rooted staggered determinant (fourth column), which are known
to deviate through $O(a^2)$ effects. Throughout, the black line is the
identity, not a fit.}
\label{fig:det}
\end{figure}

We are in an excellent position to test (\ref{detratio}), since we have all
eigenvalues with either discretization in a number of lattices with fixed
physical volume (see Tab.~\ref{tab:table1}).
Introducing
\bea
S^\mr{ov}_m(U,U')&=&\;\;\;\;\;\;\,-\log
\bigg({\det(D_m^\mr{ov}(U))\ovr\det(D_m^\mr{ov}(U'))}\bigg)
\label{def_sov}
\\
S^\mr{st}_m(U,U')&=&-{1\ovr2^{d/2}}\log
\bigg({\det(D_m^\mr{\,st}(U))\ovr\det(D_m^\mr{\,st}(U'))}\bigg)
\label{def_sst}
\eea
the goal is to show that the unfiltered ratios differ by cut-off effects only,
relation (\ref{detratio}) or
\beq
S_m^\mr{ov}(U,U')=S_m^\mr{st}(U,U')+O(a^2)
\;.
\label{logdet_nai}
\eeq
For this it is sufficient to show that an analogous relation holds
among the filtered operators
\beq
S_m^\mr{ov}(U^{(1)},U^{\prime(1)})=S_m^\mr{st}(U^{(1)},U^{\prime(1)})+O(a^2)
\;,
\label{logdet_fil}
\eeq
where the superscript $(1)$ refers to the smearing level $1$, since we already
know that the left-hand sides and the right-hand sides of (\ref{logdet_nai},
\ref{logdet_fil}) satisfy
\bea
S_m^\mr{ov}(U,U')&=&S_m^\mr{ov}(U^{(1)},U^{\prime(1)})+O(a^2)
\label{logdet_ov}
\\
S_m^\mr{st}\,(U,U')&=&S_m^\mr{st}\,(U^{(1)},U^{\prime(1)})+O(a^2)
\label{logdet_st}
\eea
respectively, because filtering amounts to an $O(a^2)$ redefinition of the
operator.

The leftmost column of Fig.~\ref{fig:det} contains scatter plots of the two
sides of (\ref{logdet_nai}) in a fixed physical volume and at a fixed quark
mass $m/e$.
Each dot represents a configuration $U$, while $U'$ is an artificial background
which realizes the ensemble mean (over $U$) of (\ref{def_sov}, \ref{def_sst}).
Obviously, the correlation on an arbitrary background $U$ gets tighter and the
pertinent slope moves closer to 1 as the lattice spacing decreases.
Note that this holds regardless of the topological charge, since our ensembles
contain a Gaussian distribution of $q$ at each $\be$.
The second column shows the scatter plots of the two sides of
(\ref{logdet_fil}), where the correlation is much better than before.
Finally, the third and fourth columns contain the scatter plots for the known
relations (\ref{logdet_ov}) and (\ref{logdet_st}).
Qualitatively, they look very similar to the first two columns, and this gives
some confidence that (\ref{logdet_nai}, \ref{detratio}) might actually
hold true.
In any case, the interesting news is that the 1-filtered staggered action
generates, on a fine enough lattice, a distribution that is closer to the
the 1-filtered overlap one, than the latter is to the unfiltered overlap
distribution.

\begin{figure}[t]
\begin{center}
\epsfig{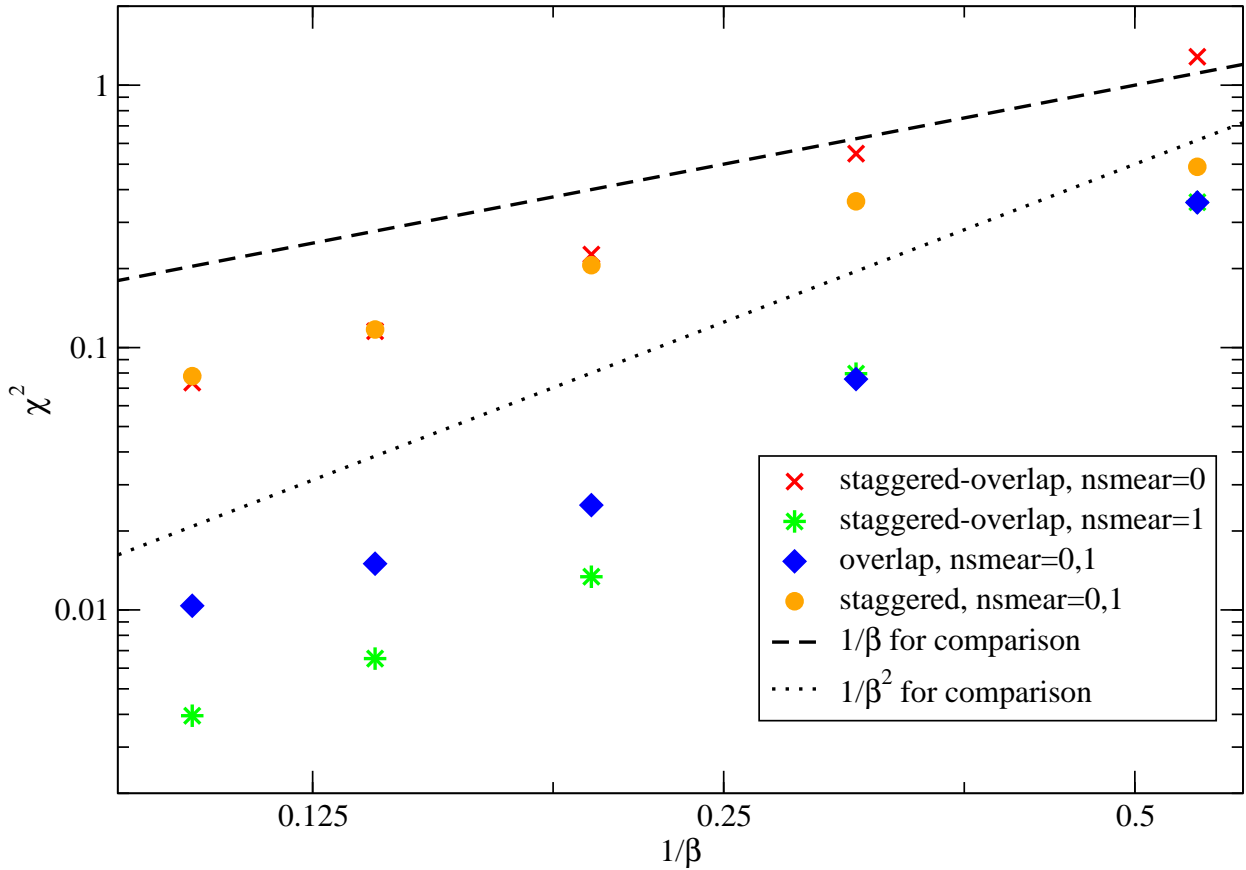}
\end{center}
\vspace{-8mm}
\caption{Scaling of $\ch^2$, defined in (\ref{def_corrcoeff}), versus
$(ae)^2$. For the diamonds/bullets it is known that they represent an
$O(a^2)$ redefinition of the operator, for the crosses/stars it is conjectured,
because of the striking similarity of the fall-off pattern.}
\label{fig:det_corrcoefficients}
\end{figure}

To make the discussion a bit more precise, we introduce a quantity designed to
measure the deviation of the correlation from the identity.
We use e.g.\ for the first column
\bea
\ch^2&=&{1\ovr \#(U)}\sum_U
\Big({1\ovr2}\log'(\det(D^\mr{st}))-\log'(\det(D^\mr{ov}))\Big)^2 
\\
\mr{where}&&
\log'(\det)=\log(\det)-{1\ovr \#(U')}\sum_{U'}\log(\det)
\nonumber
\label{def_corrcoeff}
\eea
and expect that it decreases, in a fixed physical volume, as the lattice
spacing gets smaller.
Of course, this $\ch^2$ is not a physical observable, and one should not expect
it to vanish in proportion to $1/\be\!=\!a^2$.
Still, the data in Fig.~\ref{fig:det_corrcoefficients} suggest that it falls of
--~roughly~-- with a power law in $a$.
We tested other $m/e$ which are in the scaling regime with our $\be$-values
and found qualitatively the same behavior.
The important point is that there is no obvious difference among the pattern of
the four coefficients pertinent to the four columns of Fig.~\ref{fig:det}.
Only the prefactor is different, telling us that the correlation among the
two formulations at smearing level 1 is better than the internal overlap
correlation between level 0 and 1, and the latter is better than the internal
staggered and the staggered-to-overlap correlation without smearing.
In other words, the scaling of the coefficient (\ref{def_corrcoeff}) between
the overlap and the rooted staggered contribution to the effective action
(normalized per physical volume) behaves in the same way as the one for the
internal overlap or internal rooted staggered correlation between two smearing
levels.
Since the latter are known to deviate through $O(a^2)$ effects, we rate this
as strong evidence that also the overlap and rooted staggered determinants
differ by $O(a^2)$ terms, cf.\ (\ref{logdet_nai}, \ref{logdet_fil}) or
(\ref{detratio}).


\section{Summary and discussion}


We have attempted scaling tests for dynamical overlap and rooted staggered
fermions.
The Schwinger model was chosen, because it permits to reach high numerical
accuracy while the conceptual issue of square-rooting the staggered determinant
remains for $\Nf\!=\!1$.

\bigskip

First, we have looked at the scalar condensate at fixed physical quark mass.
For $m/e\!>\!0$ the bare staggered and overlap condensates (\ref{condover},
\ref{condstag}) diverge logarithmically in $1/a$.
Subtracting the free case \emph{without\/} its non-topological zero modes
defines a UV and IR safe observable at finite quark mass, and
Fig.~\ref{fig:subt_pri_continuum} suggests that it has the same continuum value
for staggered and overlap fermions.
Hence the analytic value (\ref{schwinger}) for the condensate in the massless
1-flavor theory can be reproduced with rooted staggered fermions, if the limit
$m\!\to\!0$ is taken \emph{after\/} letting $a\!\to\!0$.
In the overlap formulation the two limits may be interchanged, while in the
staggered case the reverse order of limits yields an exact (but incorrect)
zero.
We consider the non-commutativity phenomenon (\ref{stag_noncommutativity})
important, because the Schwinger value (\ref{schwinger}) reflects the axial
anomaly, and the second version indicates that rooted staggered fermions
\emph{do get the anomaly right\/}, if the continuum limit is taken at finite
quark mass.

The second observable has been the topological susceptibility.
Both in the continuum and with overlap sea-quarks it vanishes at zero quark
mass.
Hence $\ch_\mr{top}^\mr{st}(m\!=\!0)$ is a pure lattice artefact, and it seems
to disappear like $a^2$ for an unfiltered (``naive'') staggered Dirac
operator, while in a filtered version $O(a^4...a^8)$ terms would numerically
dominate over a wide range of couplings.
On the other hand, with a finite quark mass, there is no obvious difference
among the two formulations, and we have verified that one sees pure
$O(a^2)$ scaling, as expected.
Furthermore, the continuum limit of the topological susceptibility with overlap
or rooted staggered fermions at fixed physical quark mass is consistent within
errors.

Our third observable, the partition function, is a bit more refined, but
similar in spirit, to the topological susceptibility.
Again we find, at fixed $m/e$, a continuum value, for arbitrary charge $q$,
that is consistent within errors.
Also the fourth test, a check of the Leutwyler-Smilga sum rule (\ref{LS4}) has
not revealed any difference of the two formulations in the continuum limit.

The finite temperature heavy-quark force at fixed physical distance has been our
fifth test.
It is interesting that in the Schwinger model the effects of virtual quark
loops are so pronounced and so easy to measure.
We find no deviation between the two fermion discretizations, since at filtering
level 1 overlap and rooted staggered sea-quarks agree already at the
$\be$-values considered.
It is noteworthy that the $a$-dependence of the heavy-quark selfenergy in 2D is
entirely different from the one in 4D.
As a consequence, a fuzzed heavy-quark action increases the signal-to-noise
ratio only on \emph{rough\/} lattices in 2D, while this trick helps only on
\emph{fine\/} lattices in 4D.

Finally, we have investigated the logarithmic determinant ratio of overlap and
rooted staggered fermions as a function of the lattice spacing in a fixed
physical volume. 
We find strong evidence that, at a given filtering level, this ratio is constant
over the entire configuration space up to $O(a^2)$ effects, since the pertinent
$\ch^2$ disappears in the same way as for (\ref{logdet_ov}, \ref{logdet_st})
which are established relations.
Of course, this is a numerical argument and not a proof.

In general, we find cut-off effects with dynamical overlap fermions to be
smaller than with staggered fermions.
Close to the chiral limit, the difference may be dramatic.

\bigskip

As mentioned in the introduction, the conceptual power of a scaling study is
one-sided.
We might have found an observable where the staggered answer turns out wrong in
the continuum, and this would have been sufficient to prove that the rooted
staggered action does not represent a legitimate fermion discretization.
We have not found such an observable, and the fact that even the condensate in
the 1-flavor theory is reproduced correctly (under the right order of limits)
lets one feel more sceptical whether such an observable exists.
Still, we emphasize that most of our observables involve only sea quarks, and
even the finite-mass condensate does not probe the complicated flavor/taste
structure that staggered fermions have in the valence sector.
Hence, a scaling study focusing on $\pi$ and $\et'$ properties might be
worthwhile.

For current dynamical simulations with rooted staggered quarks the implication
is two-fold.
The good news is that, for any finite sea-quark mass, we find a universal
continuum limit in a dynamical theory.
This is just numerical evidence and not a proof, but if overlap and staggered
quarks would not yield the same result in the continuum, there is no reason why
this difference would be particularly small, but non-zero, in the massive
Schwinger model.
The potentially worrisome observation is the non-commutativity phenomenon
(\ref{stag_noncommutativity}).
This is critical, because in 4D physical results are extracted by fitting the
data against predictions from staggered chiral perturbation theory (SXPT)
\cite{Aubin:2004fs}, i.e.\ the limits $a\!\to\!0$ and $m\!\to\!0$ are taken
\emph{simultaneously\/}.
It is not clear to us whether SXPT would accommodate such a non-commutativity
(in the full%
\footnote{Triggered by the current paper, Bernard has observed that SXPT
predicts similar non-commuativities in the quenched and partially quenched
theories \cite{Bernard:2004ab}.}
theory), but maybe standard physical observables in 4D QCD with $\Nf\!\geq\!2$
are not afflicted with this problem anyways.

Finally, we feel that in future investigations the focus should be on the
locality issue mentioned in the introduction.
Bunk et al.\ have shown that the fourth root (in 4D) of $D^\mr{st}$ is not a
local operator~\cite{Bunk:2004br}, but the question is, of course, whether
there is \emph{any\/} local operator which, when raised to the fourth power,
reproduces $D^\mr{st}$.
Obviously, this problem needs to be solved to lend a solid theoretical basis to
a mixed action approach as it has been explored in Ref.~\cite{Bowler:2004hs}.
Recently, Neuberger has discussed how the operation of taking the fourth root
may be cast into a local relativistic framework in 6
dimensions~\cite{Neuberger:2004be}.
The question that should be asked, in our opinion, is whether it is really
necessary to have a local operator $D$ that satisfies $D^4=D^\mr{st}$ exactly,
or whether a version with corrections, say
\beq
D^{-4}=(D^\mr{st})^{-1}\!+\!O(a^2)
\label{suggestion}
\eeq
for the Green's functions, would suffice to guarantee that a ``hybrid''
formulation with a rooted $D^\mr{st}$ in the determinant and $D$ for the
valence quarks yields the correct continuum limit%
\footnote{After the current work has been posted, three papers have appeared
that give an explicit construction of a local rooted staggered operator in the
free case \cite{Maresca:2004me,Adams:2004mf,Shamir:2004zc}. In some of them
our suggestion (\ref{suggestion}), or (6,7) in Ref.\,\cite{DuHoWe}, to allow
for cut-off effects in the interacting case has been expanded upon.}.
If so, our Fig.~\ref{fig:det} suggest that such a $D$ might be built via the
overlap prescription.

\bigskip

{\bf Acknowledgments:}
We thank Tom DeGrand and Victor Laliena for useful correspondence.
S.D.\ wishes to acknowledge discussions with Rainer Sommer, C.H.\ with
Laurent Lellouch and Leonardo Giusti.
S.D.\ was partly supported by DFG in SFB/TR-9, partly by the Swiss NSF.
C.H.\ was supported by EU grant HPMF-CT-2001-01468.




\begin{thebibliography}{99}


\itemsep -2.5pt

\bibitem{Davies:2003ik}
C.T.H.~Davies {\it et al.} [HPQCD Collaboration],
Phys.\ Rev.\ Lett.\  {\bf 92}, 022001 (2004) [hep-lat/0304004].

\bibitem{Aubin:2004fs}
C.~Aubin {\it et al.}  [MILC Collaboration],
hep-lat/0407028.

\bibitem{Bunk:2004br}
B.~Bunk, M.~Della Morte, K.~Jansen and F.~Knechtli,
Nucl.\ Phys.\ B {\bf 697}, 343 (2004) [hep-lat/0403022].

\bibitem{Hart:2004sz}
A.~Hart and E.~Muller,
hep-lat/0406030.

\bibitem{proroot_sxpt}
C.~Aubin and C.~Bernard,
Phys.\ Rev.\ D {\bf 68}, 034014 (2003) [hep-lat/0304014].
C.~Aubin and C.~Bernard,
Phys.\ Rev.\ D {\bf 68}, 074011 (2003) [hep-lat/0306026].
S.R.~Sharpe and R.S. van de Water,
hep-lat/0409018.

\bibitem{adams_onedimension}
D.H.~Adams,
Phys.\ Rev.\ Lett.\  {\bf 92}, 162002 (2004) [hep-lat/0312025].
D.H.~Adams,
hep-lat/0409013.

\bibitem{DuHo}
S.~D\"urr and C.~Hoelbling,
Phys.\ Rev.\ D {\bf 69}, 034503 (2004) [hep-lat/0311002]
and 
hep-lat/0408039.

\bibitem{DuHoWe}
S.~D\"urr, C.~Hoelbling and U.~Wenger,
Phys.\ Rev.\ D {\bf 70}, 094502 (2004) [hep-lat/0406027]
and 
hep-lat/0409108.

\bibitem{Follana:2004sz}
E.~Follana, A.~Hart and C.T.H.~Davies  [HPQCD Collaboration],
Phys.\ Rev.\ Lett.\ {\bf 93}, 241601 (2004) [hep-lat/0406010].

\bibitem{WongWolo}
K.Y.~Wong and R.M.~Woloshyn,
hep-lat/0407003.
K.Y.~Wong and R.M.~Woloshyn,
hep-lat/0412001.

\bibitem{overlap}
R.~Narayanan and H.~Neuberger,
Nucl.\ Phys.\ B {\bf 412}, 574 (1994) [hep-lat/9307006].
R.~Narayanan and H.~Neuberger,
Nucl.\ Phys.\ B {\bf 443}, 305 (1995) [hep-th/9411108].
H.~Neuberger,
Phys.\ Lett.\ B {\bf 417}, 141 (1998) [hep-lat/9707022].
H.~Neuberger,
Phys.\ Lett.\ B {\bf 427}, 353 (1998) [hep-lat/9801031].

\bibitem{Ginsparg:1981bj}
P.H.~Ginsparg and K.G.~Wilson,
Phys.\ Rev.\ D {\bf 25}, 2649 (1982).

\bibitem{Luscher:1998pq}
M.~L\"uscher,
Phys.\ Lett.\ B {\bf 428}, 342 (1998) [hep-lat/9802011].

\bibitem{Hernandez:1998et}
P.~Hernandez, K.~Jansen and M.~L\"uscher,
Nucl.\ Phys.\ B {\bf 552}, 363 (1999) [hep-lat/9808010].

\bibitem{uvfilteredstag}
T.~Blum {\it et al.},
Phys.\ Rev.\ D {\bf 55}, 1133 (1997) [hep-lat/9609036].
K.~Orginos, D.~Toussaint and R.L.~Sugar  [MILC Collaboration],
Phys.\ Rev.\ D {\bf 60}, 054503 (1999) [hep-lat/9903032].

\bibitem{Lepage:1998vj}
G.P.~Lepage,
Phys.\ Rev.\ D {\bf 59}, 074502 (1999) [hep-lat/9809157].

\bibitem{uvfilteredover}
T.~DeGrand  [MILC collaboration],
Phys.\ Rev.\ D {\bf 63}, 034503 (2001) [hep-lat/0007046].
T.~DeGrand, A.~Hasenfratz and T.G.~Kovacs,
Phys.\ Rev.\ D {\bf 67}, 054501 (2003) [hep-lat/0211006].

\bibitem{Schwinger:1962tp}
J.S.~Schwinger,
Phys.\ Rev.\ {\bf 128}, 2425 (1962).

\bibitem{Sachs:en}
I.~Sachs and A.~Wipf,
Helv.\ Phys.\ Acta {\bf 65}, 652 (1992).

\bibitem{Smilga:1995qf}
A.~Smilga and J.J.M.~Verbaarschot,
Phys.\ Rev.\ D {\bf 54}, 1087 (1996) [hep-ph/9511471].

\bibitem{MerminWagnerHohenbergColeman}
N.D.~Mermin and H.~Wagner,
Phys.\ Rev.\ Lett.\ {\bf 17}, 1133 (1966).
P.C.~Hohenberg,
Phys.\ Rev.\ 158, 383 (1967).
S.R.~Coleman,
Commun.\ Math.\ Phys.\ {\bf 31}, 259 (1973).


\bibitem{Lang:1997ib}
C.B.~Lang and T.K.~Pany,
Nucl.\ Phys.\ B {\bf 513}, 645 (1998) [hep-lat/9707024].

\bibitem{Hasenfratz:1998ri}
P.~Hasenfratz, V.~Laliena and F.~Niedermayer,
Phys.\ Lett.\ B {\bf 427}, 125 (1998) [hep-lat/9801021].

\bibitem{previousover}
R.~Narayanan, H.~Neuberger and P.~Vranas,
Nucl.\ Phys.\ Proc.\ Suppl.\ {\bf 47}, 596 (1996) [hep-lat/9509046].
R.~Narayanan, H.~Neuberger and P.~Vranas,
Phys.\ Lett.\ B {\bf 353}, 507 (1995) [hep-lat/9503013].
F.~Farchioni, I.~Hip and C.B.~Lang,
Phys.\ Lett.\ B {\bf 443}, 214 (1998) [hep-lat/9809016].
F.~Farchioni, I.~Hip, C.B.~Lang and M.~Wohlgenannt,
Nucl.\ Phys.\ B {\bf 549}, 364 (1999) [hep-lat/9812018].
S.~Chandrasekharan,
Phys.\ Rev.\ D {\bf 59}, 094502 (1999) [hep-lat/9810007].
J.E.~Kiskis and R.~Narayanan,
Phys.\ Rev.\ D {\bf 62}, 054501 (2000) [hep-lat/0001026].
L.~Giusti, C.~Hoelbling and C.~Rebbi,
Phys.\ Rev.\ D {\bf 64}, 054501 (2001) [hep-lat/0101015].

\bibitem{previousstag}
E.~Marinari, G.~Parisi and C.~Rebbi,
Nucl.\ Phys.\ B {\bf 190}, 734 (1981).
J.~Polonyi and H.W.~Wyld,
Phys.\ Rev.\ Lett.\ {\bf 51}, 2257 (1983), Erratum-ibid.\ {\bf 52}, 401 (1984).
T.~Burkitt,
Nucl.\ Phys.\ B {\bf 220}, 401 (1983).
A.N.~Burkitt and R.D.~Kenway,
Phys.\ Lett.\ B {\bf 131}, 429 (1983).
S.R.~Carson and R.D.~Kenway,
Annals Phys.\ {\bf 166}, 364 (1986).
M.~Grady,
Phys.\ Rev.\ D {\bf 35}, 1961 (1987).
V.~Azcoiti, G.~Di Carlo, A.~Galante, A.F.~Grillo and V.~Laliena,
Phys.\ Rev.\ D {\bf 50}, 6994 (1994) [hep-lat/9401032].
P.~de Forcrand, J.E.~Hetrick, T.~Takaishi and A.J.~van der Sijs,
Nucl.\ Phys.\ Proc.\ Suppl.\  {\bf 63}, 679 (1998) [hep-lat/9709104].


\bibitem{Hosotani:1997pr}
Y.~Hosotani,
hep-th/9703153.

\bibitem{Adam:1998tw}
C.~Adam,
Phys.\ Lett.\ B {\bf 440}, 117 (1998) [hep-th/9806211].




\bibitem{Leutwyler:1992yt}
H.~Leutwyler and A.~Smilga,
Phys.\ Rev.\ D {\bf 46}, 5607 (1992).


\bibitem{Colangelo:2003hf}
G.~Colangelo and S.~D\"urr,
Eur.\ Phys.\ J.\ C {\bf 33}, 543 (2004) [hep-lat/0311023].

\bibitem{Ogawa:2004ru}
K.~Ogawa and S.~Hashimoto,
hep-lat/0409103.


\bibitem{Lepage:1991ui}
G.P.~Lepage,
Nucl.\ Phys.\ Proc.\ Suppl.\  {\bf 26}, 45 (1992).

\bibitem{HQselfenergy4D}
M.~Della Morte {\it et al.} [ALPHA Collaboration],
Phys.\ Lett.\ B {\bf 581}, 93 (2004) [hep-lat/0307021].
S.~D\"urr,
hep-lat/0409141.
M.~Della Morte, F.~Knechtli, A.~Shindler, R.~Sommer,
to appear.

\bibitem{Eichten:1989zv}
E.~Eichten and B.~Hill,
Phys.\ Lett.\ B {\bf 234}, 511 (1990).


\bibitem{Bernard:2004ab}
C.~Bernard,
hep-lat/0412030.

\bibitem{Bowler:2004hs}
K.C.~Bowler, B.~Joo, R.D.~Kenway, C.M.~Maynard and R.J.~Tweedie
[UKQCD Collaboration],
hep-lat/0411005.

\bibitem{Neuberger:2004be}
H.~Neuberger,
hep-lat/0409144.

\bibitem{Maresca:2004me}
F.~Maresca and M.~Peardon,
hep-lat/0411029.

\bibitem{Adams:2004mf}
D.H.~Adams,
hep-lat/0411030.

\bibitem{Shamir:2004zc}
Y.~Shamir,
hep-lat/0412014.

\end{thebibliography}
\end{document}